%% file: Doug0725.tex
\documentclass[twocolumn]{aastex7}

\usepackage{mathtools}
\usepackage{xspace}
\usepackage{dcolumn}

\newcommand{\Ntot}{1163\xspace}
\newcommand{\Nbright}{1136\xspace}
\newcommand{\Ndwarf}{27\xspace}
\newcommand{\NComa}{41\xspace}
\newcommand{\NComaBright}{41\xspace}

\newcommand{\TFslope}{$-7.96\pm 0.13$~AB~mag\xspace}
\newcommand{\TFzpt}{$-19.34^{+0.30}_{-0.29}$~AB~mag\xspace}
\newcommand{\TFbComa}{$14.95\pm 0.02$~AB~mag\xspace}
\newcommand{\TFscatter}{$1.07\pm 0.02$~AB~mag\xspace}
\newcommand{\TFlogV}{2.01\xspace}

\shorttitle{DESI EDR Tully-Fisher Relation}
\shortauthors{Douglass et al.}

\begin{document}

\title{DESI EDR: Calibrating the Tully-Fisher Relationship with the DESI Peculiar Velocity Survey}

\input{Doug0725_author_list}

\begin{abstract}
  We calibrate the Tully-Fisher relation (TFR) with data from the DESI Peculiar 
  Velocity (PV) Survey taken during the Survey Validation (SV) period of the 
  DESI galaxy redshift survey.  Placing spectroscopic fibers on the centers and 
  major axes of spatially-extended spiral galaxies identified in the 2020 Siena 
  Galaxy Atlas using the DESI Legacy Surveys, we measure the rotational 
  velocities at 0.33$R_{26}$ for \Ntot (\Nbright + \Ndwarf dwarf) spiral 
  galaxies observed during SV.  Using \NComa spiral galaxies observed in the 
  Coma Cluster, we find a slope for the TFR of \TFslope in the $r$-band, with a 
  scatter about the TFR of \TFscatter.  We calibrate the zero-point of the TFR 
  using galaxies with independent distances measured using type Ia supernovae 
  via the cosmological distance ladder.  From the SN~Ia distances, we measure a 
  zero-point of \TFzpt in the $r$-band.  We produce a public catalog of the 
  distances to these \Nbright spiral galaxies observed during DESI SV as part of 
  the DESI PV Survey with our calibrated TFR. This is, to our knowledge, the 
  first catalog of TFR distances produced with velocities measured at a single 
  point in the disk.
\end{abstract}

\section{Introduction}

Direct measurements of individual galaxy peculiar velocities (PVs) at low 
redshift have seen a resurgence in recent years, particularly with the advent of 
large spectroscopic survey programs that can collectively observe millions of 
galaxies with high signal-to-noise \citep{York2000, Colless2001, Jones2004, 
SDSS_DR7}.  PVs have long been used as a way to constrain the growth of 
structure \citep{Kaiser1988, Gorski1989, Groth1989, Nusser1994, Davis1996, 
Park2000, Park2006, Davis2011, Johnson2014, Howlett2017c, Qin2019, Howlett2019, 
Said2020, Boruah2020, Qin2021, Turner2021, Qin2023, Lai2023, Turner2023, 
Shi2024}.  The power of PVs lies in their ability to provide a direct tracer of 
the matter field without the complications of galaxy bias, and to overcome the 
so-called ``cosmic variance limit'' that arises at low redshift when only galaxy 
redshifts are available \citep{Burkey2004, Koda2014}.  Recent studies have 
demonstrated that combined PV and redshift surveys will likely provide the 
tightest constraints on the growth of cosmic structure under the influence of 
gravity in the coming years \citep{Koda2014, Howlett2017a, Whitford2021}.

Galaxy PVs can be obtained in a number of ways, but in all cases rely on an 
empirically-calibrated redshift-independent distance indicator.  Well-known 
examples include Surface Brightness Fluctuations \citep{Tonry1988}, Type Ia 
Supernovae \citep{Phillips1993}, the Fundamental Plane relation 
\citep[FP;][]{Djorgovski1987, Dressler1987}, and the Tully-Fisher relation 
\citep[TFR;][]{Tully1977}.  Given a distance indicator, PVs can be extracted 
from the relative difference between the observed redshift and the cosmological 
(or recessional) redshift inferred from the distance indicator 
\citep{Watkins2015, Davis2014}.  The largest single samples of these various 
distance indicators now number from hundreds to several tens of thousands 
\citep{Boruah2019, Kourkchi2020a, Kourkchi2020b, Howlett2022}.  Combined, the 
largest currently available catalog is the Cosmicflows-4 compilation of 
$\sim$55,000 measurements \citep{Tully2023a}.  These data have been used to 
uncover new structures in our nearby Universe \citep{Graziani2019, Tully2023b}, 
and provide new constraints on gravity and the validity of General Relativity 
\citep{Lai2023}.  However, upcoming surveys aim to increase this catalog by a 
further order of magnitude in both cosmic volume and number of galaxies over the 
next 5 years \citep{Howlett2017b, Koribalski2020, Taylor2023}, promising further 
discoveries and tighter constraints.

One such example is the Dark Energy Spectroscopic Instrument Peculiar Velocity 
survey \citep{Saulder2023}.  Well into its 5-year program, this survey aims to 
measure, with the same instrument, PVs using both the FP and TFR methods.  
Simulations of the efficiency and success rate with which DESI can observe 
these objects predict that the full survey will contain over 130,000 and 50,000 
FP- and TFR-based PVs, respectively, over 14,000 sq. deg. (mostly in the 
northern hemisphere).  Here, we expand on the proof-of-concept provided by 
\cite{Saulder2023} by producing the first suite of DESI PV measurements using 
the now publicly available Early Data Release of DESI \citep{DESI_EDR}.  We 
present the first catalog of \Nbright TFR-based PV measurements from DESI, which 
are scattered across the full footprint but represent only a small fraction of 
the data to come. Of particular interest, and highlighted further in this work, 
are deep measurements obtained in and around the Coma Cluster.  We demonstrate 
that the DESI TFR measurements are robust, comparable to those obtained using 
\ion{H}{1} observations and other measurements \citep{Springob2007, Hong2019, Kourkchi2020b}.  The catalog is already a substantial contribution to the 
ensemble of available PV measurements, despite accounting for a small percentage 
of the data we expect with DESI's upcoming data releases.

The layout of this paper is as follows. We first introduce the Tully-Fisher 
relation and its underlying assumptions in Section~\ref{sec:TFR_background}.  We 
then describe the DESI instrument, the DESI Peculiar Velocity Survey, and the 
DESI Early Data Release (EDR) in Section~\ref{sec:DESI}.  A description of our 
Tully-Fisher measurements, our quality selection criteria, and our estimated 
systematic uncertainties follow in Section~\ref{sec:measure_rot_vel}. 
Section~\ref{sec:calibration} discusses the calibration of the slope and zero 
point of the Tully-Fisher relation using data from DESI EDR.  The measured 
Tully-Fisher relation for \Nbright galaxies in the DESI EDR is presented in 
Section~\ref{sec:measuring_pv}, followed by a comparison of our calibrated TFR 
with previous work in Section~\ref{sec:discussion}.  We conclude in 
Section~\ref{sec:conclusion}.

\section{The Tully-Fisher Relation}\label{sec:TFR_background}

Galaxy redshift surveys, by themselves, are unable to disentangle the two 
components of the observed galaxy redshift, $z_{\rm obs}$: the cosmological part 
due to the smooth expansion of the universe, $z_{\rm cosmo}$, and the peculiar 
velocity due to the gravitational attraction of the growing overdensities, 
$z_{\rm pec}$ \citep{Davis2014}:
\begin{equation}
    1 + z_\text{obs} = (1 + z_\text{cosmo})(1 + z_\text{pec})
\end{equation}
To distinguish between these two components, redshift-independent measurements 
of the distances are required.  This is possible with the use of a distance 
indicator, a correlation between distance-independent and distance-dependent 
intrinsic properties of a galaxy. 

Due to its simplicity, one of the most successful distance indicators is the 
Tully-Fisher relation \citep[TFR;][]{Tully1977}.  The TFR is a correlation 
between the rotational velocity (distance-independent property) and brightness 
(distance-dependent property) of a galaxy.  Consequently, the TFR is applicable 
to spiral galaxies, rotationally-supported systems that have an intrinsic 
relationship between their rotational velocity and luminosity. 

While there is no unanimity on the physical origins of the TFR, the 
widely-accepted model assumes the mass distribution is spherically symmetric and 
governed by Newtonian dynamics on galactic scales.  In this case, the 
centrifugal acceleration on an object moving in a circle of radius $r$ is equal 
to the gravitational acceleration on the same object attributed to the mass 
inside a sphere of radius $r$:
\begin{equation}
    V^2_{\rm rot} \propto \frac{M(r)}{r}
\end{equation}
If the mass-to-light ratio and mean surface brightness for spirals are constant, 
then the above equation can be rewritten as 
\begin{equation}\label{eqn:TFR_L}
    L \propto V_{\rm rot}^4
\end{equation}
While \cite{Aaronson1979} reproduced this relation with the infrared TFR, many 
other studies have empirically derived a power-law exponent of this relation 
deviating from 4 \citep[see, for example,][]{Sandage1976, Burstein1982, 
Bottinelli1983, Pierce1988}, which is usually attributed to the complication in 
the stellar-halo mass relation \citep{Mould2020}.

Tully-Fisher studies typically use either 21~cm \ion{H}{1} line profiles or the 
H$\alpha$ emission line to construct the maximum rotational velocities of spiral 
galaxies.  \ion{H}{1} profiles provide perhaps the most direct way of measuring 
the maximum rotational velocity of a galaxy, provided the galaxy contains a 
sufficient amount of neutral hydrogen.  At visible (rest-frame) wavelengths, the 
H$\alpha$ emission line is the most prominent and easily measurable line, making 
it the obvious choice to trace the rotational velocity of a spiral galaxy.  
However, it is more challenging to measure the maximum rotational velocity with 
H$\alpha$, since observations must be made ``far enough'' from the galactic 
center for the rotation curve to flatten \citep{Sofue2001}.  Moreover, the 
H$\alpha$ surface brightness tends to decrease more rapidly with galactocentric 
radius than \ion{H}{1} and may be weak at radii where the rotational velocity 
flattens.

Once the rotational velocity of a galaxy is measured, the TFR can be used to 
estimate the luminosity, or absolute magnitude, of the galaxy.  The absolute 
magnitude can then be compared to the apparent magnitude of the galaxy to 
determine its distance modulus.

\section{Dark Energy Spectroscopic Instrument}\label{sec:DESI}

The Dark Energy Spectroscopic Instrument \citep[DESI;][]{DESI_instrument} is a 
multi-object fiber spectrograph designed to conduct a large-scale redshift 
survey covering at least 14,000 sq.~deg.\ of the sky.  DESI's primary science 
goal is to determine the nature of dark energy by making the most precise 
measurement of the Universe's expansion history to date \citep{Levi2013}.  The 
instrument is installed on the Mayall 4-meter telescope at Kitt Peak National 
Observatory.  The robotic fibers of DESI can measure up to 5,000 spectra, 
covering a wavelength range from 3600~{\AA} to 9800~{\AA}, over 
$\sim 8~\mathrm{deg}^2$ in a single exposure \citep{DESI2016b, Silber2023, 
Miller2023, Poppett2024}, allowing the instrument to measure $\sim$40 million 
redshifts over the 5-year survey \citep{DESI2016a}.  The layout of the focal 
plane allows each fiber to move within a patrol radius that has minimal overlap 
with neighboring fibers.

Spectroscopic targets for DESI are selected from the DESI Legacy Imaging Surveys 
\citep[LS;][]{DESI_Imaging} using the DESI target selection pipeline 
\citep{DESI_target_pipeline}.  The targets are observed in tiles in a method 
following the optimized survey operations pipeline described in 
\cite{Schlafly2023}.  Redshifts are extracted in an offline spectroscopic 
pipeline using spectra measured in each fiber with the 
Redrock\footnote{\href{https://github.com/desihub/redrock}{https://github.com/desihub/redrock}.} 
template-matching software after being processed by the spectroscopic reduction 
pipeline \citep{Guy2023}.

This analysis uses spectroscopic observations made during DESI's Survey 
Validation period \citep[SV;][]{DESI_SV}, the data from which comprise the DESI 
Early Data Release \citep[EDR;][]{DESI_EDR}; to date, DESI has also released its 
first year of data \citep{DESI_DR1}.  The last phase of SV consisted of a 
reduced version of the full DESI survey, covering 1\% of the final 5-year DESI 
survey footprint.  These observations were conducted in a series of rosettes on 
the sky using the final target selection algorithms and exposures of a depth 
typical of the main survey.  For more details, see \cite{DESI_SV}.

As the main objective of DESI is to map the expansion history of the universe 
using Baryonic Accoustic Oscillations \citep[BAO;][]{DESI_DR1_BAO, 
DESI_DR2_BAO}, it has devised suitable target classes/tracers for the main 
survey: Bright Galaxies \citep[BGS;][]{DESI_target_BGS}, Luminous Red Galaxies 
\citep[LRG;][]{DESI_target_LRG}, Emission Line Galaxies 
\citep[ELG;][]{DESI_target_ELG}, and Quasi-Stellar Objects 
\citep[QSO;][]{DESI_target_QSO}, as well as Milky Way stars 
\citep[MWS;][]{DESI_target_MWS}.  In addition to these primary target classes, 
DESI also carries out various secondary targeting programs using spare fibers 
\citep{DESI_target_pipeline} to further increase the scientific value and 
targeting efficiency of the survey.  One of these programs is the DESI Peculiar 
Velocity Survey \citep{Saulder2023}.

\subsection{The DESI Peculiar Velocity Survey}

\begin{figure*}
  \centering
  \includegraphics[width=\textwidth, trim={0 0.5cm 0 0.5cm}, clip]{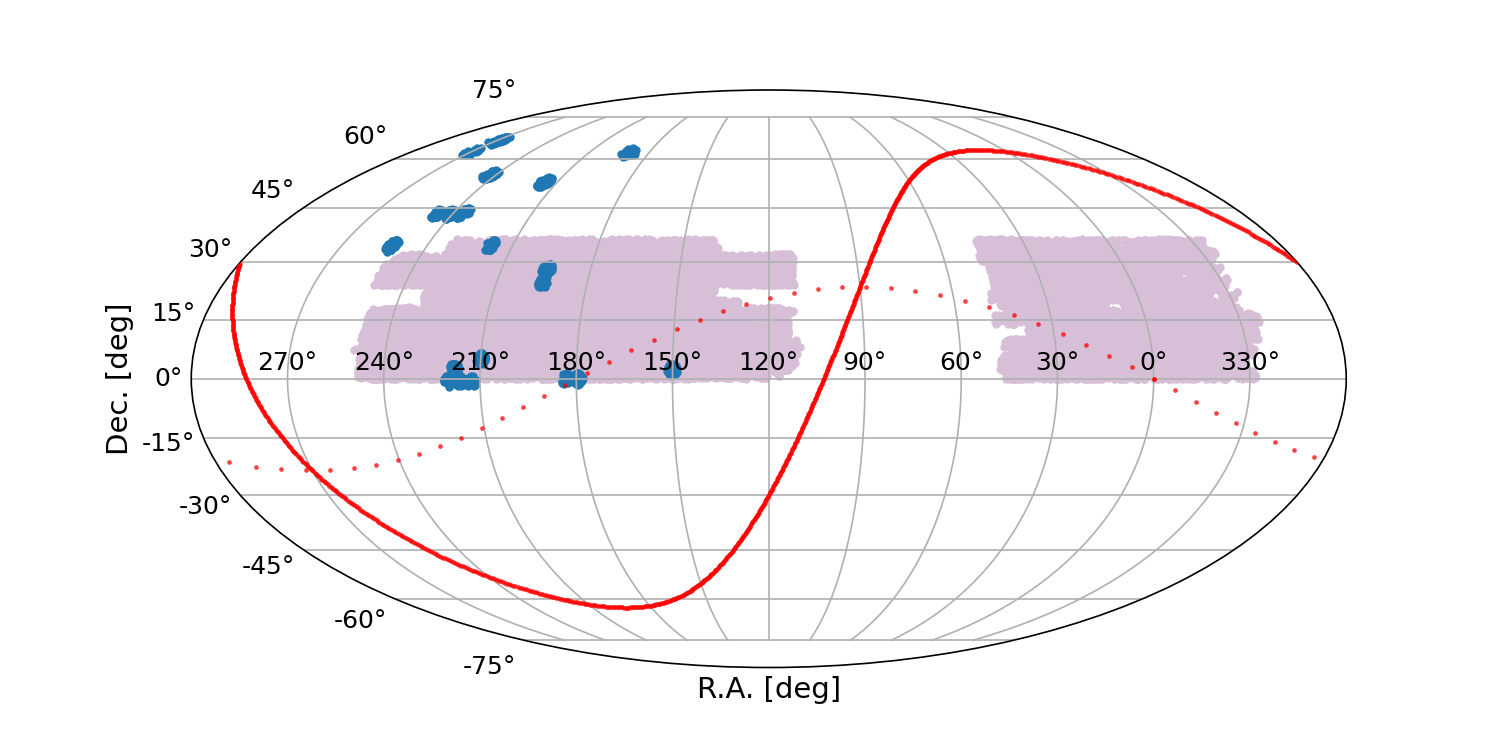}
  \caption{The distribution of Tully Fisher galaxies within the DESI EDR 
  dataset, presented here in a Mollweide projection.  The purple shaded region 
  corresponds to the ALFALFA \citep{Haynes2018} footprint, and the red solid 
  line represents the plane of the Milky Way.}
  \label{fig:sky_plot}
\end{figure*}

The DESI Peculiar Velocity (PV) Survey \citep{Saulder2023} is designed to 
measure the peculiar motions of local galaxies up to a redshift of 0.15 and 
thereby improve cosmological constraints on $f \sigma_8$, the product of the 
redshift-dependent growth rate of structure ($f$) and the amplitude of the 
linear power spectrum on the scale of 8~$h^{-1}$~Mpc ($\sigma_8$).  In addition 
to spectroscopic redshifts, redshift-independent distance indicators are also 
required to derive peculiar motions.  To this end, the DESI PV Survey targets 
suitable galaxies that can be used for either the Fundamental Plane (early-type 
galaxies) or the Tully-Fisher relation (late-type galaxies).  The target design 
of the DESI PV Survey is described in \cite{Saulder2023}.  Here, we focus on the 
Tully-Fisher sample; see \cite{Said2025} for a discussion of the Fundamental 
Plane sample and its calibration.

For the late-type galaxies used in the TFR, we select galaxies that are in the 
size-limited Siena Galaxy Atlas 2020 \citep{SGA}, built with the DESI Legacy 
Imaging Surveys DR9 \citep{DESI_Imaging} north of $-30^\circ$ declination.  The 
SGA-2020 is comprised of galaxies with diameters of the 25th magnitude 
arcsec$^{-2}$ isophote, $D(25)$, greater than 20\arcsec.  To be included in the 
TFR sample, the SGA-2020 galaxy's photometric profile must have a Sersic index 
of less than 2 \citep[characteristic of spiral galaxies;][]{Blanton2009}; and 
its inclination angle must be greater than $25^\circ$ to ensure both a 
sufficient fraction of the galaxy's rotational velocity along the line of sight 
and a confident alignment of the galaxy's position angle with its major axis.  
The distribution of TF galaxies on the sky is shown in Fig.~\ref{fig:sky_plot}.  
For each of the galaxies in the TFR sample in DESI SV, we position fibers at 
both the galactic center and at $0.33R_{26}$ on either side of the center along 
the major axis, where $R_{26}$ is the radius of the 26 mag arcsec$^{-2}$ 
$r$-band isophote.  As described in \cite{Saulder2023}, the rotational 
velocities were measured at $0.33R_{26}$ in the DESI PV Survey during DESI SV, 
as this was thought to be the furthest angular distance from the galaxies' 
centers where we could still reliably measure the redshift.  As discussed in 
\cite{Saulder2023}, the SV observations showed that we could increase the radius 
of the fiber positions slightly, so the fibers are placed at $0.4R_{26}$ in the 
DESI main survey.  

\subsection{Photometric corrections}

To adjust for various influences on the magnitudes of the galaxies, we apply a 
series of corrections to the photometry and compute a corrected apparent 
magnitude:
\begin{equation}
  m_{r, \text{corr}} = m_r - A_{\rm MW~dust} - A_{\rm internal~dust} + A_k + A_{\rm sys}
\end{equation}
where $m_r$ is the magnitude measured in the $r$-band within the 26-mag isophote 
(as given in the SGA-2020), $A_{\rm MW~dust}$ corrects for dust extinction in 
the Milky Way, $A_{\rm internal~dust}$ corrects for dust extinction internal to 
the galaxy, $A_k$ is the $k$-correction, and $A_{\rm sys}$ corrects for the 
systematic offset between the two photometric surveys used in this analysis.

We correct for the extinction due to dust in the Milky Way's interstellar medium 
(ISM) using the $E(B-V)$ map derived from $g-r$ using DESI MWS spectra 
\citep{Zhou2024}.  The magnitude correction is computed as
\begin{equation}
    A_{\rm MW~dust} = R_r E(B - V),
\end{equation}
where $R_r = 2.165$ is the ratio of total to selective extinction in the 
$r$-band through an airmass of 1.3 for a 7000~K spectrum, as discussed in \cite{Zhou2024}.

\begin{figure}
  \centering
  \includegraphics[width=0.48\textwidth]{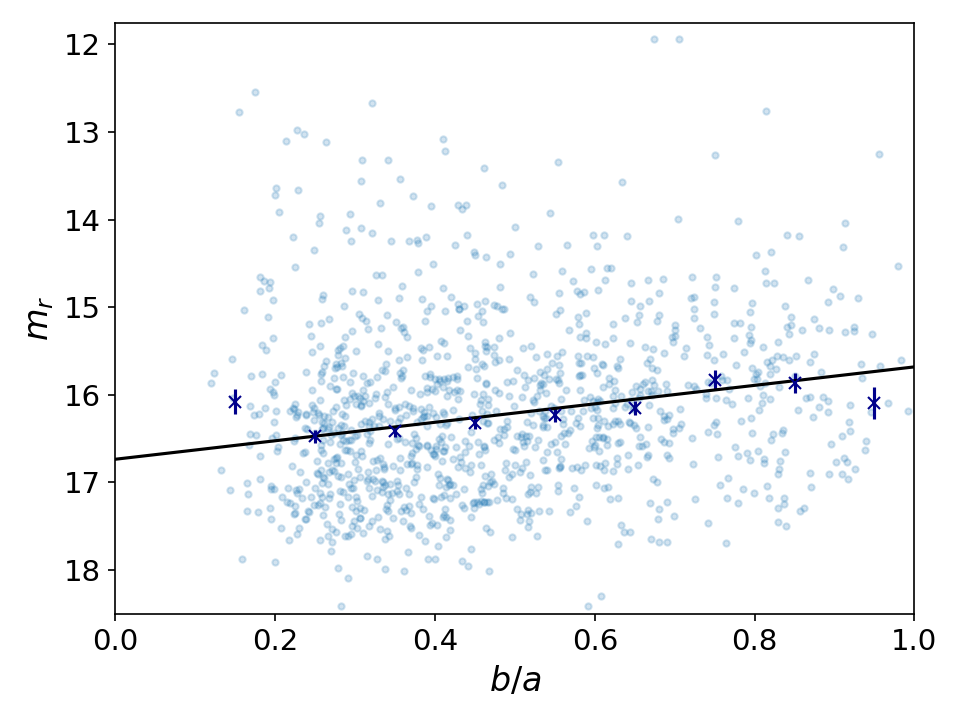}
  \caption{Observed correlation between $r$-band apparent magnitude and axis 
  ratio.  A linear fit to the median magnitudes in each bin (dark blue crosses) 
  is shown in the black line.}
  \label{fig:internal_dust_corr}
\end{figure}

Galaxies viewed at higher inclination angles will be subject to internal dust 
extinction, with the amount of extinction a function of both the galaxy's dust 
content and the inclination angle.  To correct for this extinction, we fit for 
and remove the residual correlation between the galaxy's $r$-band magnitude and 
axis ratio, shown in Fig.~\ref{fig:internal_dust_corr}.  We correct all galaxies 
so that:
\begin{equation}
    A_{\rm internal~dust} = d\left( \frac{b}{a} - 1 \right)
\end{equation}
where $d = -1.05^{+0.18}_{-0.19}$ is the slope of the line fit to the median 
magnitudes binned by the axis ratio.

The photometry used in this analysis comes from the LS, where images above Dec 
$\geq +32.375^\circ$ are from the Beijing-Arizona Sky Survey 
\citep[BASS;][]{BASS}, conducted at the Bok 2.3~m telescope at the Kitt Peak 
National Observatory (KPNO) in Arizona, and the Mayall $z$-band Legacy Survey 
\citep[MzLS;][]{DESI_Imaging}, conducted at the Mayall 4-m telescope at KPNO, 
and images below this declination were obtained using the Dark Energy Camera 
\citep[DECam;][]{DECam} at the 4~m Blanco telescope at the Cerro Tololo 
Inter-American Observatory in Chile.  Due to the use of different telescopes, 
cameras, and filters between the BASS and DECam LS (DECaLS) imaging, there are 
zero-point calibration variations between them; see Sec.~7.2 of 
\cite{DESI_Imaging} for details.  With the overlap region of the two imaging 
surveys, we can correct for these systematic variations: as described in the FP 
analysis \citep{Said2025}, it was shown that
\begin{equation}
    m_{r, \text{BASS}} - m_{r, \text{DECaLS}} = 0.0234
\end{equation}
with a root mean square (RMS) deviation of 0.02~mag and an error on the mean of 
0.0005~mag.  As in \cite{Said2025}, we adjust the northern $r$-band magnitudes 
by 0.0234 to account for this offset.

Finally, we apply $K$-corrections to the photometry with the \texttt{kcorrect} 
Python package \citep{Blanton2007} to account for observing slightly different 
parts of a galaxy's SED due to its redshift.  We $K$-correct to a redshift of 
0.05, the median redshift of our galaxy sample, to minimize this correction as 
much as possible.

\section{Measuring the rotational velocity}\label{sec:measure_rot_vel} 

\begin{figure}
  \centering
  \includegraphics[width=0.48\textwidth]{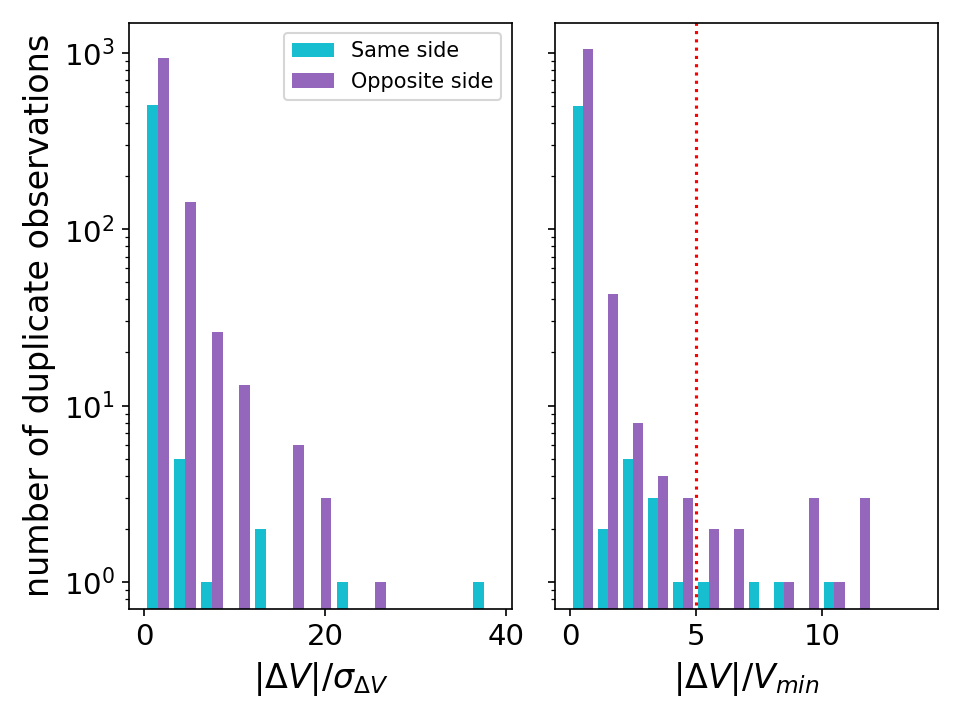}
  \caption{\emph{Left:} Pull distribution of the difference in velocity observed 
  at $0.33R_{26}$ for multiple observations on the same galaxy (on the same side 
  of the galactic center in teal, opposite sides in purple).  \emph{Right:} 
  Distribution of $|\Delta V| / V_\text{min}$ for galaxies with multiple 
  observations at $0.33R_{26}$.  All observations on a galaxy must fall to the 
  left of the dotted red line to be in the final sample.  Note that the $y$-axis 
  uses log scaling to show the tails of the distributions.}
  \label{fig:deltaV}
\end{figure}

We measure the rotational velocity of a galaxy by comparing the observed 
redshift of the galaxy's center, giving its systemic motion, and at a point 
along its major axis, which comprises both its systemic motion and the component 
of the rotational motion along the line of sight at that location in the galaxy:
\begin{equation}
  \frac{V}{c} = \frac{1 + z}{1 + z_{\rm center}} - 1
\end{equation}
To each redshift uncertainty reported by Redrock, we add in quadrature a 7~km/s 
uncertainty to account for Redrock's precision uncertainty.  As described in 
\cite{Lan2023}, this was quantified with repeat observations conducted during 
DESI SV; we use the value measured for Redrock's success on ELG spectra because 
our spectra primarily contain emission lines.   

As mentioned in Sec.~\ref{sec:TFR_background}, the TFR is frequently constructed 
using \ion{H}{1} velocities, since the slowly decreasing surface brightness of 
\ion{H}{1} as a function of galactocentric radius makes it possible to measure 
\ion{H}{1} velocities in the ``flattened'' part of the galactic rotation curve.  
However, as shown in \cite{Yegorova2007}, it is possible to calibrate the TFR at 
any galactocentric radius, so long as one is careful to only apply the TFR to 
observations made at this same radius.  The  scatter in the calibrated TFR also 
decreases with increasing radius \citep{Yegorova2007}, since more of the 
galaxies will have reached their asymptotic velocity, so it is desirable to 
calibrate with as large of a radius as possible.

In DESI observations, we apply several quality selection criteria. We require 
all galaxy center observations to have Redrock redshifts with \texttt{DELTACHI2} 
$> 25$ (the difference in the $\chi^2$ of Redrock's best-fit and second-best-fit 
spectrum model) and \texttt{ZWARN} $= 0$ (a Redrock fitting flag; $0 =$ good).  

For off-center observations, we do not require the same Redrock quality criteria 
as for the center observations because the S/N of the off-center spectra will 
necessarily be worse than the center spectra.  Instead, we require the 
calculated rotational velocity to be between 10--1000~km/s and that all 
rotational velocities with \texttt{DELTACHI2} $> 25$ at $0.33R_{26}$ satisfy 
$|\Delta V| / V_{\rm min} < 5$, where $\Delta V$ is the difference in rotational 
velocity between any two observations at that radius, and $V_\text{min}$ is the 
minimum velocity measured at that radius.  A limit of 5 on this ratio was 
selected based on visual inspection of the spectra and Redrock redshifts for 
those galaxies with multiple observations at $0.33R_{26}$.

The pull distribution for $|\Delta V|$ is shown on the left in 
Fig.~\ref{fig:deltaV}, and the distribution of $|\Delta V| / V_{\rm min}$ is 
shown on the right in Fig.~\ref{fig:deltaV}; for both, we separate the duplicate 
observations based on whether they are taken from the same side of the galaxy or 
opposite sides of the galactic center.  The uncertainty on $\Delta V$, 
$\sigma_{\Delta V}$, is the quadrature sum of the two velocities being compared: 
$\sigma_{\Delta V}^2 = \sigma_{V_1}^2 + \sigma_{V_2}^2$ for 
$\Delta V = V_1 - V_2$.  We find a wider pull distribution for observations made 
on opposite sides of the galactic center, which we primarily attribute to the 
center observation being slightly off from the kinematic center of the galaxy.  
A detailed kinematic map of the galaxy is required to correct for this potential 
systematic effect, which is unavailable to us in this analysis.  This systematic 
effect contributes to, and is therefore accounted by, the scatter in the 
calibrated TFR.

We adjust each measured velocity with the photometric inclination angle to 
estimate the total rotational velocity at that particular orbital radius in the 
galaxy:
\begin{equation}\label{eqn:Vcorrected}
  V_{\rm rot} = \frac{V}{\sin i}
\end{equation}
where
\begin{equation}\label{eqn:BAtoi}
  \cos^2 i = \frac{(b/a)^2 - q_0^2}{1 - q_0^2},
\end{equation}
$b/a$ is the photometric axis ratio from the SGA-2020, with  $q_0 = 0.2$ 
\citep{Tully2000}.

\subsection{Rotational velocity systematics}

To assess the systematics associated with measuring the rotational velocity at 
this galactocentric radius, we compare the measured rotational velocities at 
$0.33R_{26}$ in the DESI PV Survey with the expected rotational velocity at 
$0.33R_{26}$ inferred from the SDSS Mapping Nearby Galaxies at APO Survey 
\citep[MaNGA][]{MaNGA} for the same galaxies.  The comparison of the 55 galaxies 
that are in both samples is provided below.

\subsubsection{SDSS MaNGA}

\cite{Ravi2023} model the H$\alpha$ velocity maps of $\sim$5500 galaxies from 
the SDSS MaNGA DR17 \citep{SDSS_DR17}.  MaNGA used a bundle of spectroscopic 
fibers (an IFU containing between 19 and 127 fibers) to observe galaxies out to 
$1.5R_e$ or $2.5R_e$ of $\sim$10,000 galaxies \citep{Drory2015}.  The light from 
the IFUs was received by two dual-fed spectrographs covering a wavelength range 
of 3600--10,300~{\AA} with a resolution $\lambda/\Delta \lambda \sim 2000$ 
\citep{Smee2013}.

\cite{Ravi2023} fit the H$\alpha$ velocity field using the parameterization 
\begin{equation}\label{eqn:MaNGA_rot_curve}
  V(r) = \frac{V_\text{max}r}{(R_\text{turn}^\alpha + r^\alpha)^{1/\alpha}},
\end{equation}
where $V(r)$ is the tangential rotational velocity at galactocentric radius $r$, 
$V_\text{max}$ is the maximum velocity of the rotation curve, $R_\text{turn}$ is 
the radius at which the rotation curve pivots from increasing to flat, and 
$\alpha$ is a measure of how quickly the rotation curve transitions from 
increasing to flat \citep{BarreraBallesteros2018}.  During the fitting 
procedure, the following parameters are allowed to vary: $V_\text{max}$, 
$R_\text{turn}$, $\alpha$, the kinematic center of the galaxy, the inclination 
angle $i$ of the galaxy, and the position angle $\phi$ of the galaxy (the angle 
east of north of the major axis).  For more details about the modeling of the 
velocity fields, see \cite{Ravi2023}.

\subsubsection{Kinematic v. photometric rotation angle}

\begin{figure}
  \centering
  \includegraphics[width=0.48\textwidth]{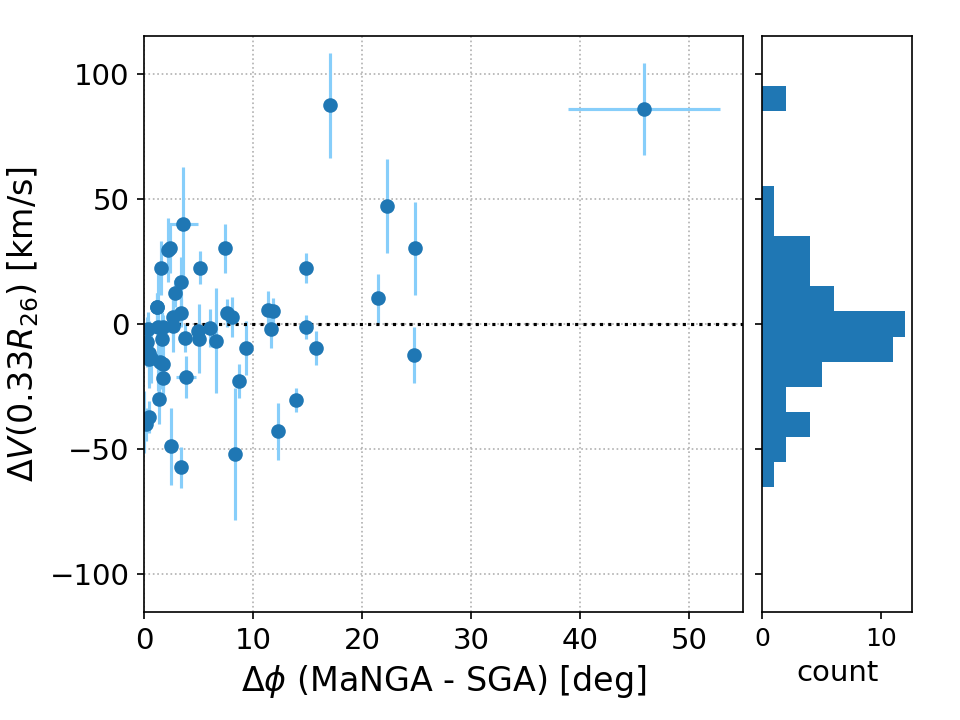}
  \caption{Difference between the velocity measured at $0.33R_{26}$ from the 
  DESI PV Survey (this work) and the rotational velocity at $0.33R_{26}$ 
  expected from the fits by \cite{Ravi2023} as a function of the difference in 
  position angle, $\Delta \phi$, between the SGA-2020 (used for targeting in the 
  DESI PV Survey) and the fitted rotation angle from \cite{Ravi2023}.  The black 
  dotted line denotes a velocity difference of zero.}
  \label{fig:dV_dPhi}
\end{figure}

We check how our rotational velocity at $0.33R_{26}$ is affected by an incorrect 
position angle for the galaxy.  As described in \cite{Saulder2023}, the DESI 
fibers are placed at $0.33R_{26}$ from the galactic center along the major axis 
of the galaxy; we assume that the galaxy's major axis is at a position angle 
$\phi$ east of north, where $\phi$ is provided by the SGA-2020 and was derived 
from photometry.  If there is a mismatch between the photometric and kinematic 
position angles, then the rotational velocity that we calculate in 
Eqn.~\ref{eqn:Vcorrected} needs an additional correction.

To test the effect of a difference between the position angles on the measured 
velocity, we compare $\Delta V (0.33 R_{26})$, the difference between the 
rotational velocity measured at $0.33R_{26}$ in the DESI PV Survey and the 
rotational velocity at the same radius predicted by the fit by \cite{Ravi2023}, 
with $\Delta \phi$, the difference between position angles from the SGA-2020 and 
that fit by \cite{Ravi2023}.  As shown in Fig.~\ref{fig:dV_dPhi}, we find no 
relationship between the velocity difference and the rotation angle offset.  The 
SDSS MaNGA DR17 H$\alpha$ velocity maps of the two outliers with 
$\Delta V(0.33R_{26})$ around 100~km/s show significant spaxel noise beyond the 
central regions of the galaxies; while the observed DESI rotational velocities 
differ greatly from that predicted by the best-fit rotation curve of 
\cite{Ravi2023}, they are consistent with the spaxel scatter at $0.33R_{26}$ for 
both of these galaxies.

\subsubsection{Scatter in DESI Rotational Velocities}\label{sec:MaNGA_Vcomp}

\begin{figure*}
  \includegraphics[width=0.49\textwidth]{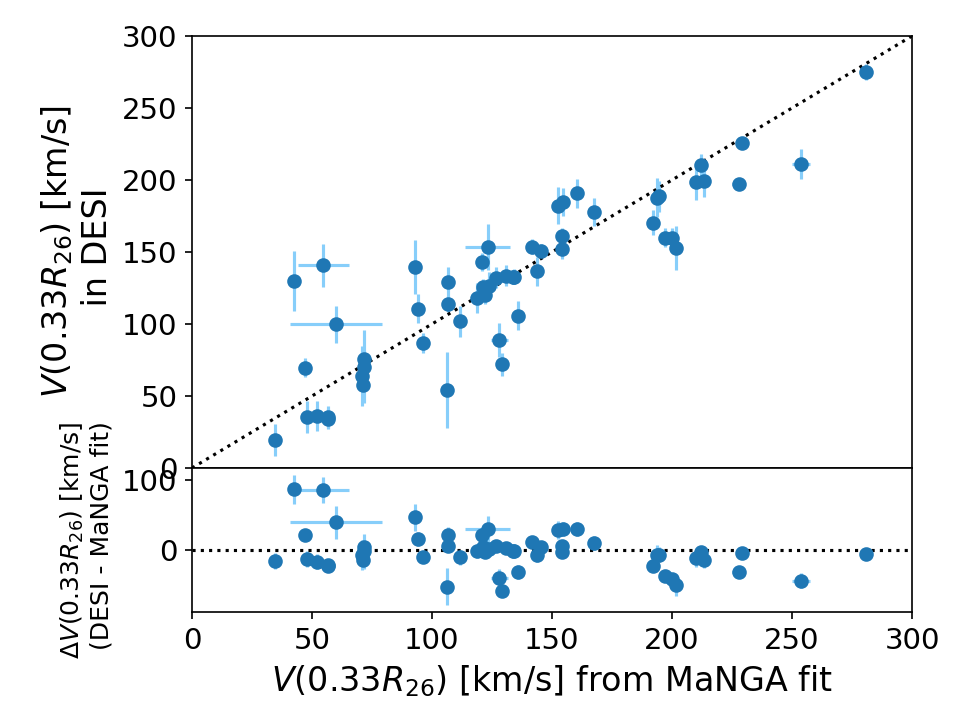}
  \includegraphics[width=0.49\textwidth]{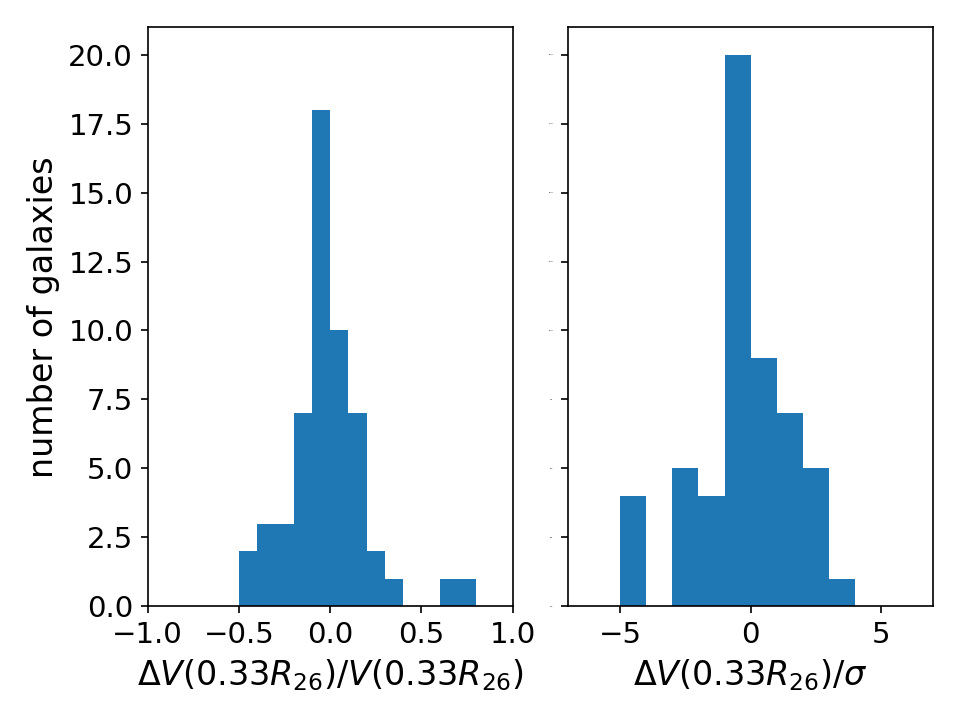}
  \caption{\emph{Left:} Comparison between the rotational velocity at 
  $0.33R_{26}$ observed in the DESI PV Survey and expected from the SDSS~MaNGA 
  DR17 rotation curve fits of \cite{Ravi2023}, corrected for differences in 
  position angle and inclination.  The black dotted line denotes equality 
  between the two velocities.  \emph{Center:} Distribution of the perpendicular 
  distance to $y = x$ normalized by the corresponding value on $y = x$ for each 
  of the galaxies appearing in the left-hand plot.  \emph{Right:} Pull 
  distribution of $\Delta V (0.33R_{26})$.}
  \label{fig:V0p33R26_MaNGA_comparison}
\end{figure*}

To quantify the scatter introduced by the measurement of the rotational velocity 
at $0.33R_{26}$, versus versus fully sampling the galaxy’s velocity field as in 
SDSS~MaNGA, we compare the DESI rotational velocity measurements at $0.33R_{26}$ 
to the predicted velocity from the rotation curve fits of \cite{Ravi2023}.

To measure just the effect that our observations have on the measured velocity, 
we convert the predicted rotational velocity from the SDSS~MaNGA fits to the 
SGA-2020 ``frame'' to account for differences in the position and inclination 
angles.  The conversion from the SDSS~MaNGA ``frame'' \citep[defined by the 
best-fit values of][]{Ravi2023} to the SGA-2020 ``frame'' is
\begin{equation}
  V_\text{MaNGA, expected} = A(Q \sin i) B(\Delta \phi) V_{\rm MaNGA},
\end{equation}
where
\begin{equation}
  A(Q \sin i) = \frac{\sin i_{\rm MaNGA}}{\sin i_{\rm SGA}},
\end{equation}
and
\begin{equation}
  B(\Delta \phi) = \cos (|\phi_{\rm MaNGA} - \phi_{\rm SGA}|).
\end{equation}
Here, quantities with a subscript ``MaNGA'' refer to the best-fit values 
reported by \cite{Ravi2023}, while quantities with the subscript ``SGA'' 
correspond to the photometric values reported in the SGA-2020.  These 
corrections transform the calculated rotational velocity so that 
$V_\text{MaNGA, expected}$ corresponds to the rotational velocity that we would 
expect to observe at $0.33R_{26}$ if the best-fit rotation curve derived by 
\cite{Ravi2023} describes the galaxy when it is oriented as defined by the 
photometric parameters of the SGA-2020; this conversion does not affect the 
uncertainty in $V_\text{MaNGA, expected}$.

The comparison of our measured velocities at $0.33R_{26}$ with those expected 
from the SDSS~MaNGA rotation curve fits is shown on the left in 
Fig.~\ref{fig:V0p33R26_MaNGA_comparison}.  The DESI observations recover the 
expected rotational velocity at $0.33R_{26}$, albeit with additional scatter.  
To quantify the scatter in this correlation, we calculate the perpendicular 
distance to the line of equality ($y = x$).  The distribution of these 
distances, normalized by the position on $y = x$ that was used to calculate this 
distance, is shown in the center panel in 
Fig.~\ref{fig:V0p33R26_MaNGA_comparison}.  The standard deviation of this 
distribution is 22\%, which will be a component of the intrinsic scatter that 
we measure in our calibrated TFR.

\section{Calibrating the Tully-Fisher Relation}\label{sec:calibration}

The TFR, shown in Eqn.~\ref{eqn:TFR_L}, is commonly written in the form
\begin{equation}\label{eqn:TFR}
  M_r = a\log \left( \frac{V}{V_0} \right) + b
\end{equation}
where $M_r$ is the absolute magnitude of the galaxy (in this work, in the 
$r$-band), $a$ is slope that corresponds to the power-law exponent of the 
relation, $b$ is the zero-point of the relation, and 
$\log (V_0 [\rm{km/s}]) =$~\TFlogV is the median $\log V$ of all galaxies used 
to calibrate the TFR.  We use a non-zero value for $\log V_0$ to shift the 
relation to minimize the correlation between the calibrated slope and 
$y$-intercepts of the TFR.  Calibrating the TFR involves determining both the 
slope and zero-point of the relation.  We follow a modification of the 
calibration scheme used in \cite{Kourkchi2020a}, where the slope is calibrated 
using galaxies in the same cluster and those with independently-measured 
distances, and the zero-point is found using the galaxies with 
independently-measured distances.

We require all galaxies used to calibrate either the slope or zero-point to have 
a minimum inclination angle of $45^{\circ}$; this is a tighter constraint than 
what we require for inclusion in the TF sample, to increase the robustness of 
our calibration.  We also visually inspect all calibration galaxies to ensure 
that there is no contamination from bright stars and/or overlapping galaxies.  
For galaxies with observed rotational velocities on both sides of the galactic 
center, we use the weighted average of the absolute value of the two velocities.

Dwarf galaxies (defined as those fainter than the line perpendicular to the 
calibrated TFR, intersecting the calibrated TFR at $M_r = -17$) are not included 
in the calibration of the TFR, as both their photometry and rotational 
velocities are typically less reliable than their brighter counterparts, and 
because they are known to have a significantly higher gas mass fraction than 
brighter galaxies.  The difference in mass fraction violates the assumption of 
the TFR that all galaxies have a uniform mass-to-light ratio.  We therefore 
remove dwarf galaxies from our TFR calibration.  However, because the slope of 
the calibrated TFR is not known until after calibration is complete, their 
removal requires an iterative process.  We first fit for the slope and 
zero-point of the TFR using all available cluster galaxies.  We then calculate 
and remove any galaxies with absolute magnitudes fainter than the dwarf limit 
based on this calibration, and we refit for the slope and zero-point of the TFR 
with this reduced galaxy sample.  We repeat this process until the galaxies 
considered to be dwarfs (i.e. below the line perpendicular to the latest 
calibration of the TFR that passes through $M_r = -17$) does not change with the 
latest calibration.  Here, we report the final values of the calibrated slope 
and zero-point after this iterative process converges.

\subsection{Defining cluster membership}\label{sec:comaclusterfinding}

\begin{figure*}
  \includegraphics[width=\textwidth]{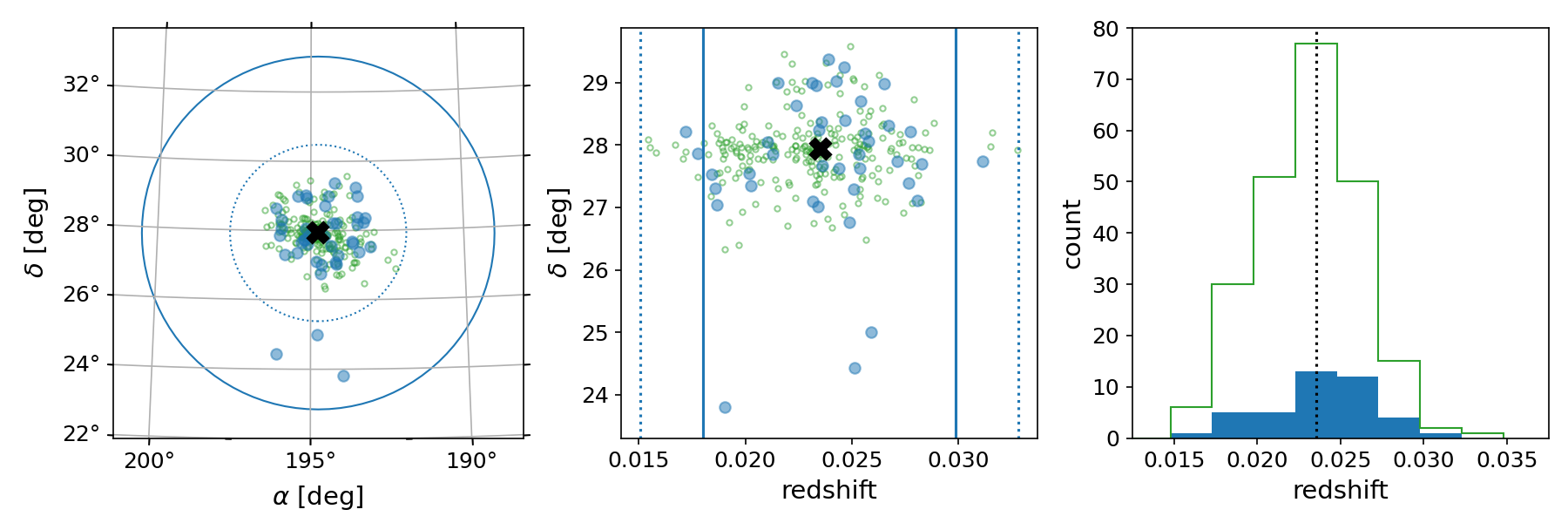}
  \caption{The distribution in $\alpha$, $\delta$, and redshift of Coma Cluster 
  galaxies used in this analysis, shown in blue.  Galaxies within the Coma 
  Cluster used in the complementary Fundamental Plane analysis by 
  \cite{Said2025} are shown in green.  The center of the Coma Cluster is denoted 
  by the black X in the first two panels and the dotted vertical line in the 
  third.  Galaxies must fall within the blue dashed lines 
  (Eqn.~\ref{eqn:cluster_membership1}) or blue solid lines 
  (Eqn.~\ref{eqn:cluster_membership2}) to be assigned to the cluster.}
  \label{fig:Coma_galaxies}
\end{figure*}

We use the Coma Cluster to calibrate the slope of the TFR for the DESI EDR, as 
it was the only cluster with a significant number of members observed during 
DESI SV.  Since the members of the Coma Cluster are all approximately the same 
distance from us, their scatter in the TFR is only due to the intrinsic scatter 
of the TFR.  Otherwise, they should all follow the same slope of the TFR for 
rotational velocities measured at the same galactocentric radius.

Similar to \cite{Kourkchi2020a}, we define galaxies in the Coma Cluster as those 
with either
\begin{align}
  R_p &< 1.5R_{2t} &\quad &\text{and} &\quad v &< V_c \pm 3\sigma_v \label{eqn:cluster_membership1}\\
  1.5R_{2t} &\leq R_p \leq 3R_{2t} &\quad &\text{and} &\quad v &< V_c \pm 2\sigma_v \label{eqn:cluster_membership2}
\end{align}
where $R_p$ is the galaxy's projected distance from the center of the Coma 
Cluster, $R_{2t} = 2.129$~Mpc \citep{Tully2015} is the Coma Cluster's projected 
second turnaround radius, $\sigma_v = 886$~km/s \citep{Tully2015} is the Coma 
Cluster's velocity dispersion (calculated from the bi-weighted line-of-sight 
velocities of its group members), and $V_c = 7176$~km/s is the average radial 
velocity of the Coma Cluster in the CMB frame \citep[computed from the cluster's 
distance modulus from][]{Tully2015}.  The limits of 
Eqn.~\ref{eqn:cluster_membership1} are depicted by the blue dotted lines in 
Fig.~\ref{fig:Coma_galaxies}, with Eqn.~\ref{eqn:cluster_membership2} shown with 
the solid blue lines.

After the inclination cut and visual inspection, we are left with \NComa 
galaxies in the Coma Cluster with which we can fit for the TFR slope.  The 
$\alpha$, $\delta$, and redshifts of these galaxies are shown in 
Fig.~\ref{fig:Coma_galaxies} and reported in Table~\ref{tab:Coma_cal}.

\input{tab1.tex}

\subsection{Zero-point calibrators}\label{sec:0pt_gals}

To calibrate the zero-point of the TFR, we use galaxies observed as part of the 
DESI PV Survey that have distances in the Extragalactic Distance Database 
\citep[EDD;][]{Tully2009}.  Of the 11 galaxies observed in the DESI PV Survey 
EDR with independent distance measurements from the EDD, only two of them 
satisfy the quality criteria described in Sec.~\ref{sec:calibration}.  Both 
their distances are calibrated from Type Ia SNe, with distance moduli measured 
by \cite{Stahl2021}.  Using the distance moduli, $\mu$, we compute the absolute 
magnitudes of these galaxies in the $r$-band:
\begin{equation}\label{eqn:absmag}
  M_r = m_r - \mu
\end{equation}
The $\alpha$, $\delta$, and redshifts of these two galaxies are reported in 
Table~\ref{tab:SNe_cal}.

\input{tab2.tex}

\subsection{Joint calibration of the slope and zero-point}\label{sec:joint_cal}

\begin{figure}
  \centering
  \includegraphics[width=0.45\textwidth]{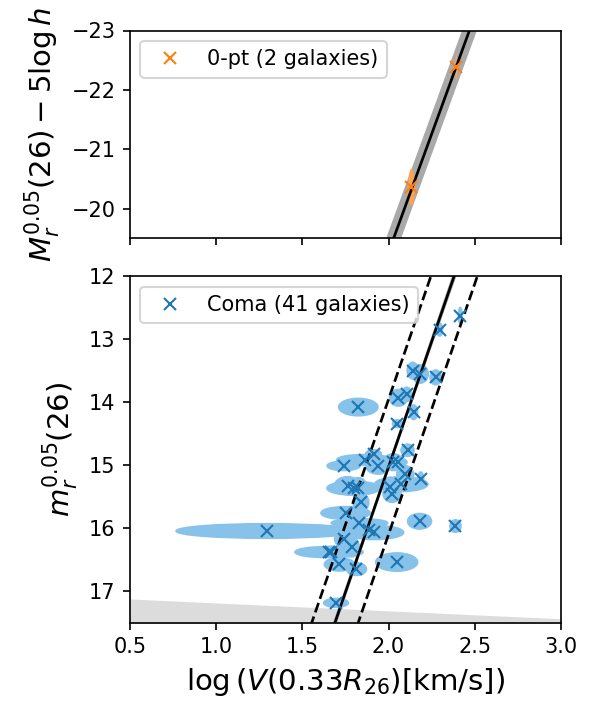}
  \caption{Calibrated TFR of the DESI EDR galaxies with $i > 45^\circ$ in the 
  Coma Cluster (lower panel; blue points) and with independent distance 
  measurements listed in the EDD (upper panel; orange points).  The shaded gray 
  region in the bottom panel denotes the region within which dwarf galaxies are 
  defined.  The calibrated TFR is drawn with the solid black line, and the 
  dispersion of the points is shown with the black dashed lines.}
  \label{fig:TFR_cal}
\end{figure}

We perform a joint fit of galaxies in the Coma Cluster and those with 
independent distances to calibrate the TFR, requiring both populations to have 
the same slope.  The fit to the population with independent distances is used to 
fix the TFR zero point, while the $y$-intercept for the Coma cluster is treated 
as a nuisance parameter.  We subclass the HyperFit\footnote{Available at 
\url{https://github.com/CullanHowlett/HyperFit}} package \citep{Robotham2015} to 
perform the fit so that we can quantify the scatter in the TFR.  HyperFit 
minimizes the distance perpendicular to the best-fit line, so we are fitting a 
hybrid of both the TFR and inverse TFR:  
\begin{equation}\label{eqn:itfr}
  a\log \left( \frac{V}{V_0} \right) = m_r - b_{\rm Coma} 
                                     = M_r - b_{\rm 0pt} 
\end{equation}
where our free parameters are $a$, $b_{\rm Coma}$, and $b_{\rm 0pt}$.  Hyperfit 
also quantifies the residual scatter along the vertical (magnitude) axis, 
$\sigma_{\rm Coma}$, based on the best fit.

As the SGA-2020 is size-limited, our calibration sample contains a 
Malmquist-like bias (with size instead of flux).  To account for this, we weight 
each galaxy in the fit by the maximum volume within which it could be included 
within the SGA-2020.  To remove any effect this normalization would have on the 
reported uncertainties, we normalize each weight by the volume within $z = 0.1$.  
Each galaxy's weight is $f_\text{max vol}^{-1}$, where
\begin{equation}
  f_\text{max vol} = \left( \frac{d_{\rm max}}{d(z = 0.1)} \right)^3
\end{equation}
where $d_{\rm max}$ is the comoving distance at which the galaxy would have an 
angular diameter of 0.2\arcmin\ (the size limit of the SGA-2020), and 
$d(z = 0.1)$ is the comoving distance at $z = 0.1$.

\begin{figure}
  \centering
  \includegraphics[width=0.49\textwidth]{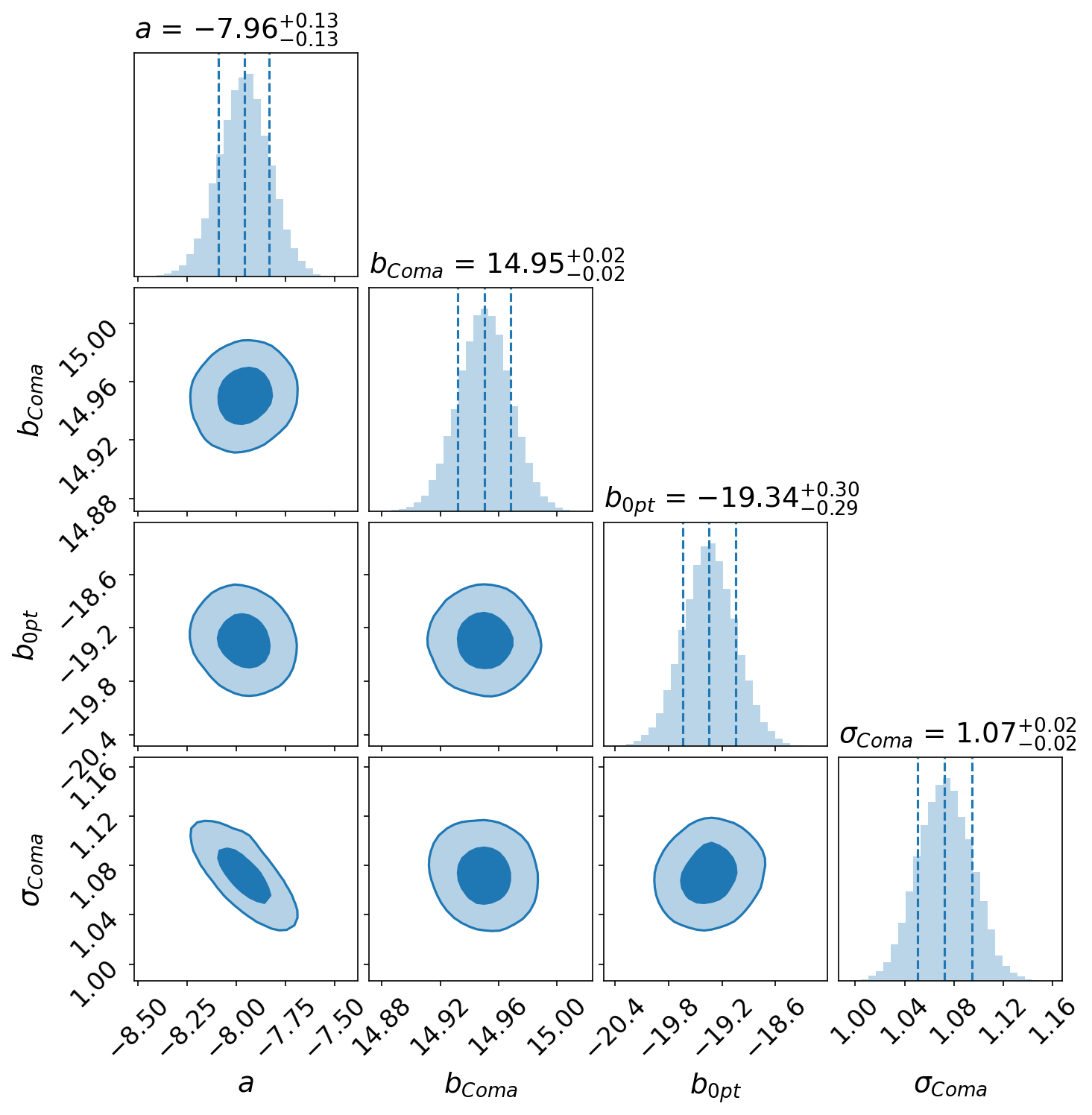}
  \caption{Corner plot of the linear fit to the Coma Cluster galaxies and 
  galaxies with independent distances shown in Fig.~\ref{fig:TFR_cal}, where $a$ 
  is the slope, $b_{\rm Coma}$ is the $y$-intercept of the Coma Cluster's TFR, 
  $b_{\rm 0pt}$ is the $y$-intercept of the galaxies with independent distances, 
  and $\sigma_{\rm Coma}$ is the intrinsic scatter in the Coma Cluster along the 
  magnitude axis.}
  \label{fig:TFR_corner}
\end{figure}

Fig.~\ref{fig:TFR_cal} shows the TFR for our galaxies in the Coma Cluster 
(bottom panel) and for the two galaxies with independent distance measures (top 
panel), with the best-fit slope value of \TFslope, a best-fit $y$-intercept 
value of \TFbComa for the Coma Cluster, and a best-fit 0-pt value of \TFzpt.  
The residual scatter along the magnitude axis for the Coma Cluster is 
\TFscatter; from our comparison to the expected velocities from SDSS~MaNGA (see 
Sec.~\ref{sec:MaNGA_Vcomp}), 0.74~AB~mag of this can be attributed to measuring 
the rotational velocity at $0.33R_{26}$.  The corner plot for the TFR 
calibration is shown in Fig.~\ref{fig:TFR_corner}.  The two galaxies with 
apparent magnitudes between 16 and 17 that have higher velocities than expected 
relative to the rest of the cluster were looked at in detail to discern why they 
might be outliers: there does not appear to be any failure in deriving the 
photometric properties in the SGA-2020, nor do their spectra look unusual, so it 
is possible that these galaxies have higher velocities than anticipated because 
their rotation curves have plateaued.

\section{Measuring Peculiar Velocities}\label{sec:measuring_pv}

\begin{figure}
  \centering
  \includegraphics[width=0.49\textwidth]{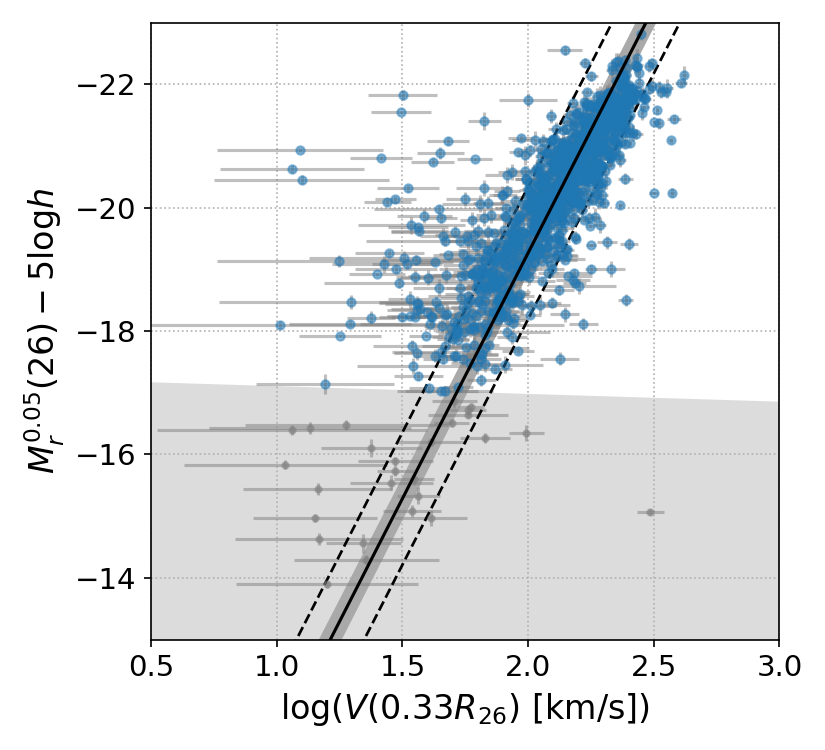}
  \caption{TFR for DESI PV Survey EDR galaxies.  Our calibrated TFR is shown in 
  the solid black line, with the light gray band representing the $1\sigma$ 
  uncertainty around the calibrated TFR; the intrinsic dispersion of the 
  calibrated TFR is shown with the black dashed lines.  The gray shaded region 
  towards the bottom denotes galaxies considered dwarfs by our calibration; 
  those that fall within it are shown as small gray points.  Vertical scatter is 
  the result of the peculiar motion of each galaxy.}
  \label{fig:TFR}
\end{figure}

\begin{figure}
  \centering
  \includegraphics[width=0.5\textwidth]{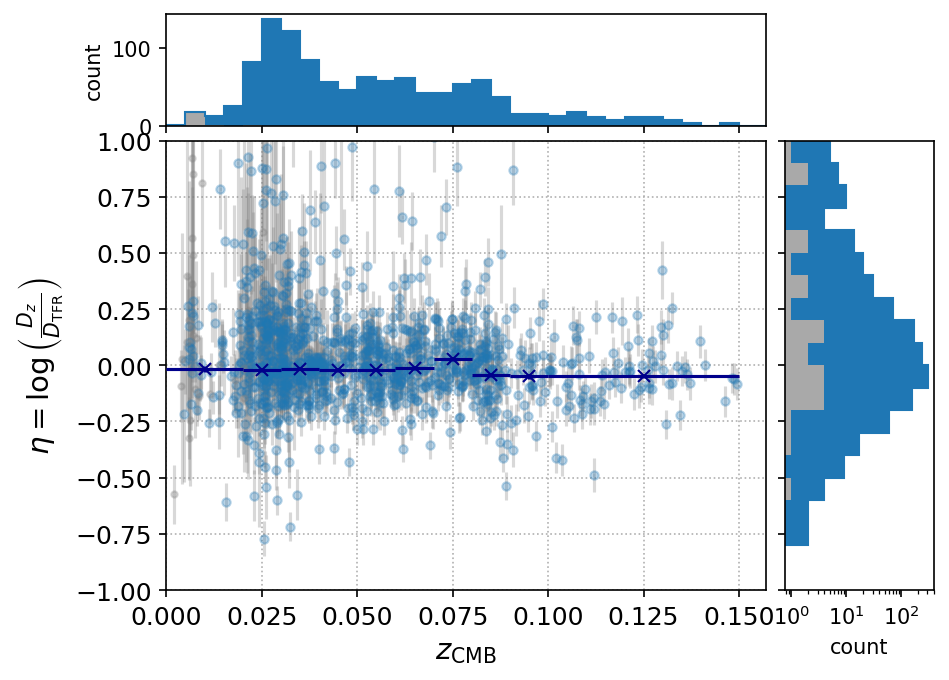}
  \caption{Log distance ratio as a function of redshift for the DESI PV Survey 
  EDR galaxies in the TF sample.  The average log distance ratio in redshift 
  bins are shown in dark blue crosses.  Galaxies considered to be dwarfs by our 
  TFR (those whose rotational velocities give an absolute magnitude below the 
  line perpendicular to the calibrated TFR at $M_r = -17$) are shown in gray.}
  \label{fig:PV_v_z}
\end{figure}

Using the calibrated slope and zero-point from Sec.~\ref{sec:calibration}, we 
calculate the distance to \Nbright spiral galaxies, out of a total of \Ntot 
spiral galaxies observed in the DESI PV Survey EDR; \Ndwarf are considered dwarf 
galaxies by our calibrated TFR, which we therefore exclude from our final 
sample.  The TFR for all of these objects is shown in Fig.~\ref{fig:TFR}, where 
the absolute magnitudes for each object are calculated using the proper distance 
for a flat $\Lambda$CDM cosmology with $\Omega_M = 0.3151$, 
$H_0 = 100h$~km/s/Mpc, and the redshift is that of the center of the galaxy as 
observed with DESI.  The vertical scatter between each galaxy and the calibrated 
TFR (shown in black) is a result of the peculiar motion of the galaxy along the 
line of sight and the intrinsic scatter of the TFR.  We calculate the peculiar 
velocity for each of the \Nbright galaxies using the \cite{Watkins2015} 
estimator.  We show the log distance ratio (the ratio of the distance to the 
galaxy measured using redshift to that from the calibrated TFR) for our sample 
as a function of redshift in Fig.~\ref{fig:PV_v_z}.  The distances measured from 
our calibrated TFR and resulting peculiar velocities for the \Nbright galaxies 
in our sample are given in Table~\ref{tab:pv}.

In both Fig.~\ref{fig:TFR} and \ref{fig:PV_v_z}, we see a significant scatter 
towards small rotational velocities and negative peculiar velocities.  After 
visual inspection, we find that the majority of galaxies more than $2\sigma$ 
from the calibrated TFR are either interacting systems, irregular galaxies, or 
relatively face-on spiral galaxies whose photometric position angles are likely 
misaligned with their kinematic position angles.  We look forward to eliminate 
some of this contamination with more accurate morphological classifications in 
future data releases, and we recommend using these galaxies' peculiar velocities 
with caution.

\input{tab3.tex}

\section{Discussion}\label{sec:discussion}
\subsection{Comparison with CosmicFlows-4}

\begin{figure}
  \centering
  \includegraphics[width=0.5\textwidth, trim={1cm 0 1cm 0}, clip]{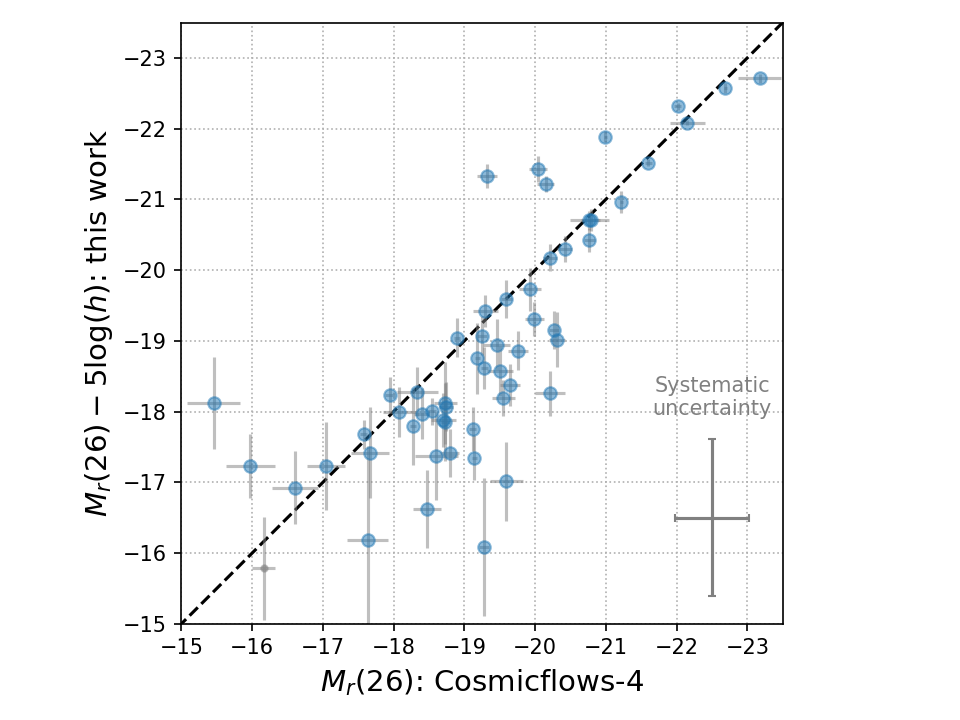}
  \caption{Comparison of the absolute magnitudes computed using our calibrated 
  TFR and that of Cosmicflows-4 \citep{Kourkchi2020a}.  Galaxies considered 
  dwarfs by our calibration ($M_r$ fainter than the line perpendicular to the 
  calibrated TFR at $M_r = -17$) are shown in small gray points.  The black 
  dashed line corresponds to $y = x$.  We find excellent agreement with their 
  TFR.}
  \label{fig:TFR_comparison}
\end{figure}

For galaxies observed as part of the DESI PV Survey EDR that have velocity 
widths observed in the ALFALFA~\ion{H}{1} survey \citep{Haynes2018}, we can 
compare our TFR calibration with the Cosmicflows-4 TFR \citep{Kourkchi2020a}.  
Because we measure the velocities at different radii and use a different value 
for $V_0$ in the calibration, it is easiest to compare our results by comparing 
the absolute magnitudes computed for the same galaxies using these two 
calibrated TFRs.  We compute the absolute magnitudes in the $r$-band with the 
distance moduli calculated by Eqn.~11 of \cite{Kourkchi2020a} using the 
parameters given in their Table~2.  We restrict our galaxies to the same domain 
within which both our calibrations are valid: spiral galaxies with 
$i > 45^\circ$, a S/N $> 10$ in the observed ALFALFA \ion{H}{1} velocity width, 
and an adjusted \ion{H}{1} velocity width $> 64$~km/s.

The comparison between our TFRs is shown in Fig.~\ref{fig:TFR_comparison} for 
the 56 galaxies that exist in both samples.  We find excellent agreement with 
the Cosmicflows-4 TFR calibration of \cite{Kourkchi2020a}.  The scatter in our 
calibrated TFR, $\sigma_{\rm Coma} =$ \TFscatter, is about twice as large as 
that of the Coma Cluster in the Cosmicflows-4 calibration \citep{Kourkchi2020a}; 
with less than half of the number of galaxies used in the calibration and 
calibrating with velocities at $0.33R_{26}$, this larger scatter is expected.  
We expect our intrinsic scatter to decrease in future DESI data releases, as we 
increase the radius at which we measure the rotational velocity to $0.4R_{26}$ 
and experiment with scaling the rotational velocities based on the surface 
brightness.

\subsection{The slope of the TFR}

As shown in Sec.~\ref{sec:joint_cal}, we measure a slope of \TFslope for the 
TFR.  In the form of Eqn.~\ref{eqn:TFR_L}, this corresponds to a power-law 
exponent of $3.18\pm0.05$.  This is smaller than the generally-accepted model 
expectation of 4, consistent with the results of \cite{Yegorova2007}, who show 
that the slope of the TFR monotonically increases with the radius at which the 
velocity is measured.  Along this same line, we should then expect our slope to 
be slightly smaller then TFR slopes calibrated with \ion{H}{1} line widths, as 
the \ion{H}{1} velocity probes the kinematics to full extent of the gaseous 
disk.  Indeed, we find that our slope is smaller than the calibrated slopes of 
the TFR for the $r$-band by, e.g., \cite{Ponomareva2017} and 
\cite{Kourkchi2020a}.

\section{Conclusion}\label{sec:conclusion}

We present a calibration of the Tully-Fisher relation (TFR) using observations 
collected in the DESI Peculiar Velocity (PV) Survey during the DESI Survey's 
Survey Validation (SV) period.  This secondary targeting program places fibers 
on the centers and major axes of spatially-resolved spiral galaxies within the 
DESI sky footprint, allowing us to measure the rotational velocities of the 
galaxies at $0.33R_{26}$ for \Ntot spiral galaxies.

We use an iterative process to calibrate the TFR, so as to perform the 
calibration without dwarf galaxies.  We fit for the TFR using galaxies in the 
Coma Cluster, the only cluster observed during DESI SV with a high enough galaxy 
count to perform this calibration, and two galaxies observed in DESI SV with 
calibrated distances from SN~Ia, with four free parameters: the slope, 
$y$-intercepts of the Coma Cluster and galaxies with known distances, and the 
intrinsic scatter in the Coma Cluster's TFR.  We then calculate the line 
perpendicular to the Coma Cluster's TFR that intersects the cluster's TFR at an 
apparent magnitude corresponding to $M_r = -17$ at the distance of the Coma 
Cluster, and we remove any galaxies within the Coma Cluster that are fainter 
than this line.  We repeat this process (fitting for the slope and 
$y$-intercepts) until we converge on a fixed set of non-dwarf galaxies in Coma.  
This results in \NComaBright galaxies in Coma used to fit for a slope of 
\TFslope in the $r$-band for velocities at $0.33R_{26}$.  The calibrated zero 
point is \TFzpt.

This work serves as a proof-of-concept for the analysis using the full 5-year 
DESI PV Survey.  The substantial increase in sample size will greatly reduce the 
uncertainties in our calibrated Tully-Fisher relation and will permit the 
measure of the peculiar motion of more than 53,000 disk galaxies 
\citep{Saulder2023} in the local universe, composing the largest sample of disk 
galaxies with measured peculiar velocities to date.

\begin{acknowledgements}

The authors thank former NSF REU undergraduate students Grace Chiodo, for her 
help with the SDSS~MaNGA comparisons, and Hayley Nofi, for her help implementing 
the subclassed version of Hyperfit.  This material is based on work supported in 
part by the National Science Foundation Grant No. PHY-1757062 and PHY-2149332 
and the Department of Energy Grant No. SC0008475.  Data shown in figures, as 
well as example Python code to generate the figures, are available on Zenodo at 
\url{https://doi.org/10.5281/zenodo.15794231}.

This material is based upon work supported by the U.S. Department of Energy 
(DOE), Office of Science, Office of High-Energy Physics, under Contract No. 
DE–AC02–05CH11231, and by the National Energy Research Scientific Computing 
Center, a DOE Office of Science User Facility under the same contract.  
Additional support for DESI was provided by the U.S. National Science Foundation 
(NSF), Division of Astronomical Sciences under Contract No. AST-0950945 to the 
NSF’s National Optical-Infrared Astronomy Research Laboratory; the Science and 
Technology Facilities Council of the United Kingdom; the Gordon and Betty Moore 
Foundation; the Heising-Simons Foundation; the French Alternative Energies and 
Atomic Energy Commission (CEA); the National Council of Humanities, Science and 
Technology of Mexico (CONAHCYT); the Ministry of Science, Innovation and 
Universities of Spain (MICIU/AEI/10.13039/501100011033), and by the DESI Member 
Institutions: \url{https://www.desi.lbl.gov/collaborating-institutions}.  Any 
opinions, findings, and conclusions or recommendations expressed in this 
material are those of the author(s) and do not necessarily reflect the views of 
the U.S. National Science Foundation, the U.S. Department of Energy, or any of 
the listed funding agencies.

The authors are honored to be permitted to conduct scientific research on 
I'oligam Du'ag (Kitt Peak), a mountain with particular significance to the 
Tohono O’odham Nation.

\end{acknowledgements}

\software{
Astropy \citep{astropy:2013, astropy:2018, astropy:2022}, 
Corner \citep{corner}, 
HyperFit \citep{Robotham2015}, 
Kcorrect \citep{Blanton2007},
Matplotlib \citep{matplotlib},
NumPy \citep{numpy}, 
SciPy \citep{scipy}
}

\bibliographystyle{aasjournal}
\bibliography{Doug0725_sources}
\end{document}

%% file: Doug0725_author_list.tex
\author[0000-0002-9540-546X]{K.~Douglass}
\affiliation{Department of Physics \& Astronomy, University of Rochester, 500 Joseph C. Wilson Blvd., Rochester, NY 14627, USA}
\email[show]{kellyadouglass@rochester.edu}

\author[0000-0001-5537-4710]{S.~BenZvi}
\affiliation{Department of Physics \& Astronomy, University of Rochester, 500 Joseph C. Wilson Blvd., Rochester, NY 14627, USA}
\email{sybenzvi@pas.rochester.edu}

\author[0000-0002-7517-9629]{N.~Uberoi}
\affiliation{Physics Department, Yale University, P.O. Box 208120, New Haven, CT 06511, USA}
\email{navya.uberoi@yale.edu}

\author[0000-0002-1081-9410]{C.~Howlett}
\affiliation{School of Mathematics and Physics, University of Queensland, Brisbane, QLD 4072, Australia}
\email{c.howlett@uq.edu.au}

\author[0000-0002-0408-5633]{C.~Saulder}
\affiliation{Max Planck Institute for Extraterrestrial Physics, Gie\ss enbachstra\ss e 1, 85748 Garching, Germany}
\email{csaulder@mpe.mpg.de}

\author[0000-0002-1809-6325]{K.~Said}
\affiliation{School of Mathematics and Physics, University of Queensland, Brisbane, QLD 4072, Australia}
\email{k.saidahmedsoliman@uq.edu.au}

\author[0000-0002-7852-167X]{R.~Demina}
\affiliation{Department of Physics \& Astronomy, University of Rochester, 500 Joseph C. Wilson Blvd., Rochester, NY 14627, USA}
\email{regina@pas.rochester.edu}

\author{J.~Aguilar}
\affiliation{Lawrence Berkeley National Laboratory, 1 Cyclotron Road, Berkeley, CA 94720, USA}
\email{jaguilar@lbl.gov}

\author[0000-0001-6098-7247]{S.~Ahlen}
\affiliation{Department of Physics, Boston University, 590 Commonwealth Avenue, Boston, MA 02215 USA}
\email{ahlen@bu.edu}

\author{G.~Aldering}
\affiliation{Lawrence Berkeley National Laboratory, 1 Cyclotron Road, Berkeley, CA 94720, USA}
\email{galdering@lbl.gov}

\author[0000-0001-9712-0006]{D.~Bianchi}
\affiliation{Dipartimento di Fisica ``Aldo Pontremoli'', Universit\`{a} degli Studi di Milano, Via Celoria 16, I-20133 Milano, Italy}
\affiliation{INAF-Osservatorio Astronomico di Brera, Via Brera 28, 20122 Milano, Italy}
\email{davide.bianchi1@unimi.it}

\author{D.~Brooks}
\affiliation{Department of Physics \& Astronomy, University College London, Gower Street, London, WC1E 6BT, UK}
\email{david.brooks@ucl.ac.uk}

\author{T.~Claybaugh}
\affiliation{Lawrence Berkeley National Laboratory, 1 Cyclotron Road, Berkeley, CA 94720, USA}
\email{tmclaybaugh@lbl.gov}

\author[0000-0002-2169-0595]{A.~Cuceu}
\affiliation{Lawrence Berkeley National Laboratory, 1 Cyclotron Road, Berkeley, CA 94720, USA}
\email{acuceu@lbl.gov}

\author[0000-0002-4213-8783]{T.~M.~Davis}
\affiliation{School of Mathematics and Physics, University of Queensland, Brisbane, QLD 4072, Australia}
\email{tamarad@physics.uq.edu.au}

\author[0000-0002-0553-3805]{K.~S.~Dawson}
\affiliation{Department of Physics and Astronomy, The University of Utah, 115 South 1400 East, Salt Lake City, UT 84112, USA}
\email{kdawson@astro.utah.edu}

\author[0000-0002-1769-1640]{A.~de la Macorra}
\affiliation{Instituto de F\'{\i}sica, Universidad Nacional Aut\'{o}noma de M\'{e}xico,  Circuito de la Investigaci\'{o}n Cient\'{\i}fica, Ciudad Universitaria, Cd. de M\'{e}xico  C.~P.~04510,  M\'{e}xico}
\email{macorra@fisica.unam.mx}

\author[0000-0002-3033-7312]{A.~Font-Ribera}
\affiliation{Institut de F\'{i}sica d’Altes Energies (IFAE), The Barcelona Institute of Science and Technology, Edifici Cn, Campus UAB, 08193, Bellaterra (Barcelona), Spain}
\email{afont@ifae.es}

\author[0000-0002-2890-3725]{J.~E.~Forero-Romero}
\affiliation{Departamento de F\'{i}sica, Universidad de los Andes, Cra. 1 No. 18A-10, Edificio Ip, CP 111711, Bogot\'{a}, Colombia}
\affiliation{Observatorio Astron\'{o}mico, Universidad de los Andes, Cra. 1 No. 18A-10, Edificio H, CP 111711 Bogot\'{a}, Colombia}
\email{je.forero@uniandes.edu.co}

\author{E.~Gazta\~{n}aga}
\affiliation{Institut d'Estudis Espacials de Catalunya (IEEC), c/ Esteve Terradas 1, Edifici RDIT, Campus PMT-UPC, 08860 Castelldefels, Spain}
\affiliation{Institute of Cosmology and Gravitation, University of Portsmouth, Dennis Sciama Building, Portsmouth, PO1 3FX, UK}
\affiliation{Institute of Space Sciences, ICE-CSIC, Campus UAB, Carrer de Can Magrans s/n, 08913 Bellaterra, Barcelona, Spain}
\email{enrique.gaztanaga@port.ac.uk}

\author[0000-0003-3142-233X]{S.~Gontcho A Gontcho}
\affiliation{Lawrence Berkeley National Laboratory, 1 Cyclotron Road, Berkeley, CA 94720, USA}
\affiliation{Department of Astronomy, University of Virginia, Charlottesville, VA 22904, USA}
\email{satyagontcho@lbl.gov}

\author{G.~Gutierrez}
\affiliation{Fermi National Accelerator Laboratory, PO Box 500, Batavia, IL 60510, USA}
\email{gaston@fnal.gov}

\author[0000-0003-1197-0902]{C.~Hahn}
\affiliation{Steward Observatory, University of Arizona, 933 N. Cherry Avenue, Tucson, AZ 85721, USA}
\email{chhahn@arizona.edu}

\author[0000-0002-6550-2023]{K.~Honscheid}
\affiliation{Center for Cosmology and AstroParticle Physics, The Ohio State University, 191 West Woodruff Avenue, Columbus, OH 43210, USA}
\affiliation{Department of Physics, The Ohio State University, 191 West Woodruff Avenue, Columbus, OH 43210, USA}
\email{kh@physics.osu.edu}

\author[0000-0002-6024-466X]{M.~Ishak}
\affiliation{Department of Physics, The University of Texas at Dallas, 800 W. Campbell Rd., Richardson, TX 75080, USA}
\email{mishak@utdallas.edu}

\author[0000-0003-0201-5241]{R.~Joyce}
\affiliation{NSF NOIRLab, 950 N. Cherry Ave., Tucson, AZ 85719, USA}
\email{richard.joyce@noirlab.edu}

\author[0000-0003-3510-7134]{T.~Kisner}
\affiliation{Lawrence Berkeley National Laboratory, 1 Cyclotron Road, Berkeley, CA 94720, USA}
\email{tskisner@lbl.gov}

\author[0000-0001-6356-7424]{A.~Kremin}
\affiliation{Lawrence Berkeley National Laboratory, 1 Cyclotron Road, Berkeley, CA 94720, USA}
\email{akremin@lbl.gov}

\author[0000-0003-1838-8528]{M.~Landriau}
\affiliation{Lawrence Berkeley National Laboratory, 1 Cyclotron Road, Berkeley, CA 94720, USA}
\email{mlandriau@lbl.gov}

\author[0000-0003-1887-1018]{M.~E.~Levi}
\affiliation{Lawrence Berkeley National Laboratory, 1 Cyclotron Road, Berkeley, CA 94720, USA}
\email{melevi@lbl.gov}

\author[0000-0002-9748-961X]{J.~Lucey}
\affiliation{Institute for Computational Cosmology, Department of Physics, Durham University, South Road, Durham DH1 3LE, UK}
\email{john.lucey@durham.ac.uk}

\author[0000-0002-4279-4182]{P.~Martini}
\affiliation{Center for Cosmology and AstroParticle Physics, The Ohio State University, 191 West Woodruff Avenue, Columbus, OH 43210, USA}
\affiliation{Department of Astronomy, The Ohio State University, 4055 McPherson Laboratory, 140 W 18th Avenue, Columbus, OH 43210, USA}
\email{martini.10@osu.edu}

\author[0000-0002-1125-7384]{A.~Meisner}
\affiliation{NSF NOIRLab, 950 N. Cherry Ave., Tucson, AZ 85719, USA}
\email{aaron.meisner@noirlab.edu}

\author{R.~Miquel}
\affiliation{Instituci\'{o} Catalana de Recerca i Estudis Avan\c{c}ats, Passeig de Llu\'{\i}s Companys, 23, 08010 Barcelona, Spain}
\affiliation{Institut de F\'{i}sica d’Altes Energies (IFAE), The Barcelona Institute of Science and Technology, Edifici Cn, Campus UAB, 08193, Bellaterra (Barcelona), Spain}
\email{rmiquel@ifae.es}

\author[0000-0002-2733-4559]{J.~Moustakas}
\affiliation{Department of Physics and Astronomy, Siena College, 515 Loudon Road, Loudonville, NY 12211, USA}
\email{jmoustakas@siena.edu}

\author[0000-0003-3188-784X]{N.~Palanque-Delabrouille}
\affiliation{IRFU, CEA, Universit\'{e} Paris-Saclay, F-91191 Gif-sur-Yvette, France}
\affiliation{Lawrence Berkeley National Laboratory, 1 Cyclotron Road, Berkeley, CA 94720, USA}
\email{nathalie.palanque-delabrouille@cea.fr}

\author[0000-0002-0644-5727]{W.~J.~Percival}
\affiliation{Department of Physics and Astronomy, University of Waterloo, 200 University Ave W, Waterloo, ON N2L 3G1, Canada}
\affiliation{Perimeter Institute for Theoretical Physics, 31 Caroline St. North, Waterloo, ON N2L 2Y5, Canada}
\affiliation{Waterloo Centre for Astrophysics, University of Waterloo, 200 University Ave W, Waterloo, ON N2L 3G1, Canada}
\email{will.percival@uwaterloo.ca}

\author[0000-0001-7145-8674]{F.~Prada}
\affiliation{Instituto de Astrof\'{i}sica de Andaluc\'{i}a (CSIC), Glorieta de la Astronom\'{i}a, s/n, E-18008 Granada, Spain}
\email{f.prada@csic.es}

\author{G.~Rossi}
\affiliation{Department of Physics and Astronomy, Sejong University, 209 Neungdong-ro, Gwangjin-gu, Seoul 05006, Republic of Korea}
\email{graziano@sejong.ac.kr}

\author[0000-0002-9646-8198]{E.~Sanchez}
\affiliation{CIEMAT, Avenida Complutense 40, E-28040 Madrid, Spain}
\email{eusebio.sanchez@ciemat.es}

\author{D.~Schlegel}
\affiliation{Lawrence Berkeley National Laboratory, 1 Cyclotron Road, Berkeley, CA 94720, USA}
\email{djschlegel@lbl.gov}

\author{M.~Schubnell}
\affiliation{Department of Physics, University of Michigan, 450 Church Street, Ann Arbor, MI 48109, USA}
\email{schubnel@umich.edu}

\author[0000-0002-3461-0320]{J.~Silber}
\affiliation{Lawrence Berkeley National Laboratory, 1 Cyclotron Road, Berkeley, CA 94720, USA}
\email{jhsilber@lbl.gov}

\author{D.~Sprayberry}
\affiliation{NSF NOIRLab, 950 N. Cherry Ave., Tucson, AZ 85719, USA}
\email{david.sprayberry@noirlab.edu}

\author[0000-0003-1704-0781]{G.~Tarl\'{e}}
\affiliation{Department of Physics, University of Michigan, 450 Church Street, Ann Arbor, MI 48109, USA}
\email{gtarle@umich.edu}

\author{B.~A.~Weaver}
\affiliation{NSF NOIRLab, 950 N. Cherry Ave., Tucson, AZ 85719, USA}
\email{benjamin.weaver@noirlab.edu}

\author[0000-0001-5381-4372]{R.~Zhou}
\affiliation{Lawrence Berkeley National Laboratory, 1 Cyclotron Road, Berkeley, CA 94720, USA}
\email{rongpuzhou@lbl.gov}

\author[0000-0002-6684-3997]{H.~Zou}
\affiliation{National Astronomical Observatories, Chinese Academy of Sciences, A20 Datun Road, Chaoyang District, Beijing, 100101, P.~R.~China}
\email{zouhu@nao.cas.cn}

%% file: tab1.tex
\begin{deluxetable*}{cccCcCC}
\tablewidth{0pt}
\tablehead{\colhead{SGA-2020 ID} & \colhead{R.A.} & \colhead{Decl.} & \colhead{Redshift} & \colhead{$D(26)$} & \colhead{$m_r(26)$} & \colhead{$V(0.33R_{26})$} 
\\[-0.5em] 
& [deg] & [deg] & & [arcmin] & [\text{AB mag}] & [\text{km/s}]
}
\tablecaption{Coma cluster galaxies used for TFR slope calibration\label{tab:Coma_cal}}
\startdata
25532  & 195.976671 & 28.310624 & 0.02673(2) & 0.79 & 16.91\pm0.006 & \phantom{0}46.5\pm\phantom{0}7.2\\
98934  & 194.206914 & 27.093896 & 0.02315(1) & 1.40 & 14.38\pm0.011 & 188\phantom{.0}\pm13\phantom{.0} \\
122260 & 195.847744 & 27.306878 & 0.01857(2) & 0.45 & 17.30\pm0.018 & \phantom{0}65.2\pm\phantom{0}8.7 \\
191275 & 194.575941 & 27.848483 & 0.02534(4) & 0.68 & 16.74\pm0.026 & \phantom{0}20\phantom{.0}\pm24\phantom{.0} \\
196592 & 193.638307 & 27.632752 & 0.02439(2) & 1.49 & 14.45\pm0.023 & 153\phantom{.0}\pm16\phantom{.0} \\
202666 & 195.212967 & 27.742984 & 0.02712(5) & 0.76 & 16.44\pm0.011 & \phantom{0}56\phantom{.0}\pm19\phantom{.0} \\
221178 & 193.067420 & 27.533159 & 0.01843(1) & 1.02 & 15.24\pm0.005 & \phantom{0}81.8\pm\phantom{0}8.3 \\
291879 & 195.433257 & 28.999533 & 0.02153(3) & 1.40 & 14.73\pm0.018 & 114\phantom{.0}\pm13\phantom{.0} \\
309306 & 193.526516 & 29.242659 & 0.02465(5) & 0.97 & 16.03\pm0.016 & 119\phantom{.0}\pm42\phantom{.0} \\
337817 & 194.168156 & 28.217658 & 0.02773(2) & 0.61 & 16.64\pm0.037 & 151\phantom{.0}\pm24\phantom{.0} \\
364410 & 193.281401 & 28.241242 & 0.02346(2) & 0.63 & 17.01\pm0.020 & \phantom{0}55.5\pm\phantom{0}7.1 \\
364929 & 194.213359 & 29.378171 & 0.02391(4) & 1.38 & 14.98\pm0.005 & \phantom{0}67\phantom{.0}\pm17\phantom{.0} \\
365429 & 194.181327 & 27.034746 & 0.01869(3) & 0.70 & 15.46\pm0.029 & \phantom{0}73\phantom{.0}\pm27\phantom{.0} \\
366393 & 194.661188 & 27.013210 & 0.02340(2) & 0.69 & 16.87\pm0.013 & \phantom{0}77.0\pm\phantom{0}5.9 \\
378842 & 194.270257 & 28.209628 & 0.01718(2) & 0.43 & 17.64\pm0.006 & \phantom{0}49.6\pm\phantom{0}8.0 \\
455486 & 196.088608 & 24.430244 & 0.02514(2) & 1.12 & 16.28\pm0.027 & \phantom{0}66.9\pm\phantom{0}7.3 \\
465951 & 196.148924 & 28.627694 & 0.02239(2) & 0.62 & 15.48\pm0.006 & \phantom{0}55\phantom{.0}\pm12\phantom{.0} \\
479267 & 193.223264 & 28.371257 & 0.02358(2) & 2.09 & 13.43\pm0.012 & 260.6\pm\phantom{0}5.4 \\
486394 & 193.975798 & 23.802604 & 0.01904(2) & 0.81 & 16.48\pm0.031 & \phantom{0}69.1\pm\phantom{0}7.0 \\
566771 & 194.666055 & 26.759415 & 0.02486(2) & 1.05 & 16.21\pm0.021 & \phantom{0}57.9\pm\phantom{0}9.9 \\
645151 & 195.130309 & 28.950469 & 0.02334(2) & 1.10 & 16.31\pm0.028 & 102.3\pm\phantom{0}7.2 \\
747077 & 194.538607 & 28.708611 & 0.02542(2) & 2.30 & 14.35\pm0.014 & 138.1\pm\phantom{0}9.8 \\
748600 & 193.655970 & 27.691999 & 0.02825(3) & 0.62 & 15.77\pm0.048 & 107\phantom{.0}\pm14\phantom{.0} \\
753474 & 195.164866 & 29.019325 & 0.02427(3) & 2.38 & 13.52\pm0.006 & 197.7\pm\phantom{0}7.4 \\
819754 & 194.119091 & 27.291300 & 0.02508(2) & 1.13 & 14.77\pm0.005 & 112.5\pm\phantom{0}6.9 \\
826543 & 195.188419 & 27.747018 & 0.03111(8) & 0.68 & 17.47\pm0.013 & 112\phantom{.0}\pm31\phantom{.0} \\
841705 & 195.960413 & 28.054375 & 0.02102(3) & 0.70 & 15.81\pm0.032 & \phantom{0}87\phantom{.0}\pm14\phantom{.0} \\
917608 & 193.444254 & 27.385853 & 0.02765(1) & 0.77 & 16.35\pm0.007 & 105\phantom{.0}\pm12\phantom{.0} \\
995924 & 194.348213 & 27.549951 & 0.02019(6) & 0.46 & 16.96\pm0.016 & \phantom{0}45\phantom{.0}\pm20\phantom{.0} \\
1050173 & 195.038059 & 27.866433 & 0.01775(2) & 0.52 & 15.38\pm0.016 & 114\phantom{.0}\pm13\phantom{.0} \\
1167691 & 193.464618 & 28.979424 & 0.02649(2) & 1.10 & 14.86\pm0.017 & 139.9\pm\phantom{0}6.9 \\
1203610 & 195.313399 & 27.669224 & 0.02360(3) & 0.78 & 16.09\pm0.024 & \phantom{0}63\phantom{.0}\pm22\phantom{.0} \\
1203786 & 194.431835 & 29.003335 & 0.02313(2) & 0.72 & 17.18\pm0.038 & \phantom{0}61.0\pm\phantom{0}6.2 \\
1269260 & 194.802578 & 25.003968 & 0.02589(3) & 0.53 & 17.18\pm0.014 & \phantom{0}52\phantom{.0}\pm10\phantom{.0} \\
1274409 & 193.474729 & 28.186669 & 0.02561(2) & 0.88 & 15.39\pm0.017 & 129\phantom{.0}\pm11\phantom{.0}\\
1284002 & 193.480767 & 28.393845 & 0.02467(3) & 0.97 & 16.62\pm0.008 & 244\phantom{.0}\pm18\phantom{.0} \\
1323268 & 195.158001 & 28.057463 & 0.02578(2) & 1.11 & 14.65\pm0.008 & 127.3\pm\phantom{0}7.7 \\
1356626 & 195.093287 & 27.623562 & 0.02537(4) & 0.52 & 16.36\pm0.028 & \phantom{0}68\phantom{.0}\pm25\phantom{.0} \\
1364394 & 194.819484 & 27.106093 & 0.02804(1) & 0.52 & 15.95\pm0.014 & 153.7\pm\phantom{0}5.4 \\
1379275 & 195.449906 & 27.354792 & 0.02026(5) & 0.68 & 16.82\pm0.009 & \phantom{0}83\phantom{.0}\pm32\phantom{.0} \\
1387991 & 196.019819 & 27.850501 & 0.02129(3) & 0.73 & 15.97\pm0.034 & 125\phantom{.0}\pm14\phantom{.0}
\enddata
\tablecomments{{List} of the \NComaBright galaxies in the Coma cluster used for calibrating the slope of the TFR.  Sky positions and diameters of the 26 mag arcsec$^{-2}$ isophote in the $r$-band are from the SGA-2020 \citep{SGA}.  Redshifts are measured from the DESI EDR spectra, and rotational velocities at $0.33R_{26}$ are computed as described in Sec.~\ref{sec:measure_rot_vel}.}
\end{deluxetable*}

%% file: tab2.tex
\begin{deluxetable*}{cCCCcCCCc}
\tablewidth{0pt}
\tablehead{\colhead{SGA-2020 ID} & \colhead{R.A.} & \colhead{Decl.} & \colhead{Redshift} & \colhead{$D(26)$} & \colhead{$m_r(26)$} & \colhead{$V(0.33R_{26})$} & \colhead{$\mu$} & \colhead{SN} 
\\[-0.5em] 
& [\text{deg}] & [\text{deg}] &  & \colhead{[arcmin]} & [\text{AB mag}] & \colhead{[\text{km/s}]} & [\text{mag}] & }
\tablecaption{Galaxies used for TFR zero-point calibration\label{tab:SNe_cal}}
\startdata
294387 & 183.547194 & -0.831611 & 0.02486(2) & 1.22 & 14.95\pm0.005 & 135.7\pm6.0 & 34.77\pm0.30 & SN2019kcx \\
464075 & 242.480987 & 43.129126 & 0.03283(3) & 1.88 & 14.06\pm0.016 & 247.4\pm3.6 & 35.56\pm0.16 & 2006cc
\enddata
\tablecomments{{List} of the two galaxies used for calibrating the zero-point of the TFR.  Sky positions and diameters of the 26 mag arcsec$^{-2}$ isophote in the $r$-band are from the SGA-2020 \citep{SGA}.  Redshifts are measured from the DESI EDR spectra, and rotational velocities at $0.33R_{26}$ are computed as described in Sec.~\ref{sec:measure_rot_vel}.  Distance moduli are from \cite{Stahl2021}.}
\end{deluxetable*}

%% file: tab3.tex
\begin{deluxetable*}{cCCCcCCCC}
\tablewidth{0pt}
\tablehead{\colhead{SGA-2020 ID} & \colhead{R.A.} & \colhead{Decl.} & \colhead{Redshift} & \colhead{$D(26)$} & \colhead{$m_r(26)$} & \colhead{$V(0.33R_{26})$} & \colhead{$\mu$} & \colhead{$V_{\rm pec}$} 
\\[-0.5em] 
& [\text{deg}] & [\text{deg}] & & [arcmin] & [\text{AB mag}] & [\text{km/s}] & [\text{AB mag}] & [\text{km/s}]}
\tablecaption{DESI EDR Peculiar Velocities from the TFR\label{tab:pv}}
\startdata
896  & 180.381827 & \phantom{0}1.295827 & 0.02243(3) & 0.62 & 15.92\pm0.005 & \phantom{0}64\phantom{.0}\pm15\phantom{.0} & 33.32\pm0.92 & \phantom{-1}3100\pm3000 \\
1548 & 252.860800 & 33.034304 & 0.06211(2) & 0.54 & 17.07\pm0.010 & 145\phantom{.0}\pm10\phantom{.0} & 37.01\pm0.39 & \phantom{1}-4700\pm3300 \\
1583 & 236.811826 & 43.995408 & 0.03547(2) & 0.44 & 17.31\pm0.016 & \phantom{0}64.8\pm\phantom{0}6.2 & 34.65\pm0.46 & \phantom{-1}2700\pm2200 \\
1980 & 221.076523 & \phantom{0}0.044044 & 0.02825(2) & 1.05 & 14.66\pm0.009 & 316\phantom{.0}\pm12\phantom{.0} & 37.75\pm0.32 & -11900\pm1300 \\
2497 & 240.338401 & 42.491513 & 0.04404(2) & 0.50 & 15.61\pm0.011 & 197.1\pm\phantom{0}5.6 & 36.90\pm0.32 & \phantom{1}-7400\pm1900
\enddata
\tablecomments{{Five} of the \Nbright galaxies in DESI EDR with peculiar velocities measured using the calibrated TFR.  Sky positions and diameters of the 26 mag arcsec$^{-2}$ isophote in the $r$-band are from the SGA-2020 \citep{SGA}.  Redshifts are measured from the DESI EDR spectra, and rotational velocities at $0.33R_{26}$ are computed as described in Sec.~\ref{sec:measure_rot_vel}.  Distance moduli are calculated from the calibrated TFR, and peculiar velocities are based on the difference between the observed redshift and that inferred from the distance moduli following \cite{Watkins2015}.  Table~\ref{tab:pv} is published in its entirety online in a machine-readable format.  A portion is shown here for guidance regarding its form and content.}
\end{deluxetable*}